\documentclass[]{aastex631}

\usepackage{graphicx}
\usepackage[utf8]{inputenc}
\usepackage[english]{babel}
\usepackage{natbib}
\usepackage{fp}
\usepackage[capbesideposition=right]{floatrow}
\usepackage{wrapfig}
\usepackage{mathtools}
\usepackage{amsmath}
\usepackage{supertabular,enumitem}
\usepackage{tabularx}
\usepackage{float}

\accepted{July 22, 2024}

\shorttitle{Connecting the Low to High Corona}
\shortauthors{Alzate et al.}

\graphicspath{{./}{figures/}}

\begin{document}


\title{Connecting the Low to High Corona:  Propagating Disturbances as Tracers of the Near-Sun Solar Wind}

\correspondingauthor{Nathalia Alzate}
\email{nathalia.alzate@nasa.gov}

\author[0000-0001-5207-9628]{Nathalia Alzate}
\affiliation{NASA Goddard Space Flight Center \\
Greenbelt MD 20771, USA}
\affiliation{ADNET Systems, Inc. \\
Greenbelt MD 20771, USA}

\author[0000-0001-6407-7574]{Simone Di Matteo}
\affiliation{NASA Goddard Space Flight Center \\
Greenbelt MD 20771, USA}
\affiliation{The Catholic University of America  \\
Washington, DC, USA}

\author[0000-0002-6547-5838]{Huw Morgan}
\affiliation{Aberystwyth University \\
Ceredigion, Cymru SY23 3BZ, UK}

\author[0000-0003-1692-1704]{Nicholeen Viall}
\affiliation{NASA Goddard Space Flight Center \\
Greenbelt MD 20771, USA}

\author[0000-0002-8164-5948]{Angelos Vourlidas}
\affiliation{Johns Hopkins University Applied Physics Laboratory \\
Laurel, MD 20723, USA}



\begin{abstract}

We revisit a quiet 14-day period of solar minimum during January 2008 and track sub-streamer propagating disturbances (PDs) from low heights in STEREO/EUVI to the extended corona through STEREO/COR1 and into STEREO/COR2 along nonradial paths that trace the structure of the underlying streamers.  Using our recently developed method for generating nonradial Height-Time profiles of outward PDs (OPDs) and inward PDs (IPDs), we obtained their velocities along the radial and position angle directions.  Our analysis of 417 unique OPDs revealed two classes: slow and fast OPDs. Slow OPDs form preferentially at $\approx$1.6 $R_\odot$ closer to the streamer boundaries, with asymmetric occurrence rates, and show speeds of $16.4_{-8.4}^{+26.6}km/s$ at 1.5 $R_\odot$ and accelerate up to $200.1_{-57.9}^{+71.1}km/s$ at 7.5 $R_\odot$. Fast OPDs form preferentially at $\approx$ 1.6 $R_\odot$ and at $\approx$3.0 $R_\odot$ both at the streamer boundaries and slightly more often within them. They show speeds of $87.8_{-24.8}^{+59.1}km/s$  at 1.5 $R_\odot$ up to $197.8_{-46.7}^{+61.8}km/s$ at 7.5 $R_\odot$. IPDs are observed forming at $\approx$1.8 $R_\odot$ with speeds of tens of $km/s$, mostly concentrated in the aftermath of a CME eruption. We present an example in which we show that periodic brightness variations related to OPDs remained in the range of 98 to 128 min, down to $\approx$2.0 $R_\odot$, well within the field of view of COR1. The velocity profiles of slow OPDs for heliocentric height below 3.0 $R_\odot$ show good agreement with speeds more closely related to the bulk solar wind obtained via interplanetary scintillation.


\end{abstract}




\section{Introduction} 
\label{sec:intro}

To understand the origin of the slow solar wind, attention must be paid to the dynamic behavior of density structures and waves low in the corona and down to the lower layers of the solar atmosphere. Many variations visible in the low and middle corona accelerate with the solar wind and maintain coherence at least through $\sim$20 $R_{\odot}$ \citep{DeForest2016}, thus can be used as solar wind tracers. This indicates that below this height, the variations visible in solar images are likely structures frozen into the solar wind, and not turbulent fluctuations.  Furthermore, remote sensing and \textit{in situ} data strongly suggest that much of the structure observed in the slow wind is a tracer of its formation.  However, as of yet, there is no fully-understood link between observed features in the corona and \textit{in situ}-detected mesoscale structures in the heliosphere \citep{Viall2021}.  Establishing a relationship between these structures and the origin of the slow wind is required to improve solar wind models \citep{Viall2020}.  

Multi-scale magnetic reconnection in the corona has been theorized to cause the variability in the slow wind.  Close to the solar surface, jetlets \citep{Raouafi2014, Raouafi2016, Raouafi2023} have been suggested as the source of ``microstreams'' and were found to correspond to structured solar wind observed \textit{in situ} by Parker Solar Probe (PSP) \citep{Fox2016,Kumar2022, Kumar2023}.  Other studies favor reconnection events on a larger scale, namely events near streamer cusps and global null points \citep{Suess1996, Einaudi1999, Wang2000}.  Evidence of this structured solar wind is further supported by observations of transients in PSP/WISPR \citep{Howard2019,Poirier2023, Reville2020} and SolO/Metis \citep{Antonucci2020,Ventura2023} remote sensing observations.  In this context, the S-web concept relates the solar wind's origin to reconnection occurring between closed and open field lines \citep{Fisk2001, Antiochos2011, Zhao2011, Wang2012, Edmondson2012, Higginson2017}.  The slow solar wind has thus been associated with different sources near coronal hole/streamer boundaries, an indication that it may form through the continuous release of density structures \citep{Antiochos2011, Einaudi2001, Lapenta2005}.  Indeed, recent studies have described pseudostreamers and null point topologies (examples of S-web arc) as events in which reconnection and plasma release into the heliosphere occur \citep{Mason2019, Stansby2018, DiMatteo2019}.  

Streamers show considerable temporal variation of density and are host to several small-scale dynamic features \citep{Sheeley1997, DeForest2018} that have been observed to propagate from the corona into the solar wind.  One example of small-scale, non-turbulent structures is helmet streamer plasmoids, or blobs commonly known as ``Sheeley blobs'', which are structures that place constraints on the acceleration and source of the slow solar wind.  \citet{Sheeley1997, Sheeley2007, Sheeley2009, Wang1998, Wang2000} and \citet{Harrison2009} observed these structures in white light images, and \citet{Crooker2004, Suess2009, Sheeley2010} used composition/magnetic field data to indirectly connect helmet streamer plasmoids to 1 AU plasmoids.  \citet{Rouillard2009, Rouillard2010, Rouillard2011} tracked these plasmoids to 1 AU and together with \citet{Suess2009}, and \citet{Sanchez2017a}, they established that ``Sheeley blobs'' are created through magnetic reconnection at the tips of helmet streamers.

Another example of small transient structures are the smaller-scale periodic density structures, or PDSs, on radial size scales of 70-3000 Mm and exhibiting characteristic periodicities ranging from a few minutes to a few hours.  These structures have been detected by \textit{in situ} measurements \citep{Kepko2016} and linked to the density structures released into the solar wind \citep{Viall2008, Viall2009SK, Viall2010}.  PDSs occur as short as $\sim$5 minutes and form at or below $\sim$2.5 $R_{\odot}$ \citep{Viall2010, Viall2015}.  Often, they survive to L1 \citep{Viall2008, Viall2009, Kepko2024} where they can directly drive the Earth's magnetosphere in a ``breathing mode'' \citep{Kepko2002, Kepko2003, Viall2009, Claudepierre2009, Hartinger2014} and affect radiation belt electrons \citep{Kepko2019,DiMatteo2022}.  The PDSs accelerate with the slow wind through the region between $\sim$2.5 and 84 $R_{\odot}$.  Their source below this region, however, has yet to be determined.  \citet{Kepko2024} provided various scenarios in which magnetic reconnection is a key aspect in the mechanism by which PDSs are created.  

If observations of small-scale transients above $\sim$2.5 $R_{\odot}$ and \textit{in situ} measurements are indicative of magnetic reconnection occurring near the sun surface as the mechanism by which the solar wind forms, it stands to reason that a signature would be left behind in remote sensing observations below $\sim$2.5 $R_{\odot}$.  Indeed, \citet{Wang1999, Sheeley2001, Sheeley2002, Sheeley2007, Sheeley2014}, and \citet{Sanchez2017a} have identified reconnection events in the region between 2.0 and 5.0 $R_{\odot}$.  \citet{Hess2017} identified reconnection events at distances below $\sim$2.0 $R_{\odot}$.  These studies describe downward-moving density enhancements, or inflows, associated with outward-moving plasma, or blobs.  \citet{Sanchez2017a} also made a direct connection between the two and thus concluded that the mechanism by which inflows and blobs form is the same.  A study by \cite{Seaton2021} identified inflows that appear to arise from reconnection related to streamer detachments that subsequently appear to disturb the lower corona.  

Implications for the sources of the slow solar wind can be made by identifying density structures in the low corona from the field of view (FOV) of EUV imagers (up to $\sim$1.7 $R_{\odot}$) and inner coronagraphs (between $\sim$1.4 and $\sim$ 4.0 $R_{\odot}$), observing the features that give rise to them, and connecting them to the high corona in the FOV of outer coronagraphs (up to $\sim$15 $R_{\odot}$).  However, the complexity of the plasma configuration at the base of the corona hinders the process of directly connecting the slow solar wind to small-scale time-varying structures observed in the corona below $\sim$2.5 $R_{\odot}$.  Plasma configurations are shaped by the coronal magnetic field, which exhibits a nonradial nature starting at the sun until approximately $\sim$2.5-3.0 $R_{\odot}$ \citep{Boe2020}.  Additionally, connecting the myriad of structures in this region to structures higher up has proven difficult due to data limitations in terms of noise reduction.  \citet{Alzate2021} developed a method that suppresses both high- and low-frequency variations in STEREO/SECCHI observations.  Their work presented evidence of tracers of the solar wind in the low corona EUV and WL observations near and at the streamer location.  Additionally, \citet{Alzate2023} described a method for tracking nonradial outflows in EUV and WL images.  Their methodology made a more robust case on the reliability of the results of \citet{Alzate2021}.  In this paper, we revisit the 14-day period of low coronal activity described in \citet{Alzate2021,Alzate2023} and present an in-depth analysis of the outflow events identified.  This last work opens the door to a more reliable analysis of the kinematics of these features including nonradial motion in the plane-of-sky (POS) that was not resolved in the original work \citep{Alzate2021} and is more relevant in the low corona.  In Section 2 we describe the data and methods used in this new analysis.  Section 3 presents our results, which we discuss in Section 4.  We present our conclusions in Section 5.

\section{Data and Methods} 
\label{sec:datameth}

\subsection{Datasets and Image Processing}
\label{sec:imgproc}

We used data from the Sun Earth Connection and Heliospheric Investigations \citep[SECCHI,][]{Howard2008} suite of instruments on board the Solar Terrestrial Relations Observatory Ahead and Behind \citep[STEREO-A and -B,][]{Kaiser2008} twin spacecraft.  Specifically, we used observations by the Extreme Ultraviolet Imager \citep[EUVI,][]{Wuelser2004} in the 195 \AA\ channel, the COR1 inner coronagraph \citep{Thompson2003}, and the COR2 outer coronagraph.  The EUVI instrument observes the sun and the corona out to $\sim$1.7 $R_{\odot}$, COR1 observes between 1.4 and 4.0 $R_{\odot}$, and COR2 observes between 2.5 and 15 $R_{\odot}$.  Here and throughout this paper, we refer to heliocentric heights. Together, the three instruments offer an uninterrupted view of the corona.  As for \citet{Alzate2021,Alzate2023}, the data used in this study are from the STEREO-A spacecraft where the streamer we are revisiting is on the eastern limb and an additional streamer analyzed for the first time in this paper is on the western limb.  

Following the steps described in \citet{Alzate2021}, both EUV and WL images were processed using the Bandpass Frame Filtering (BFF) method. The core of the processing method is a temporal bandpass filter that effectively damps high-frequency noise and low-frequency slow-changing disturbances.  The filtering is achieved through convolution with two normalized kernels defined as a wide and a narrow Gaussian kernel (see \citet{Alzate2021} for details).  For this study, we tuned the filters to isolate components on timescales between $\approx$1.25 and $\approx$10.3 hours. For the images needed in the nonradial method described below (see Section \ref{sec:htt-plots}), we applied the Normalizing Radial Graded Filter (NRGF) method \citep{Morgan2006}, which is a simple spatial filter for removing the steep radial gradient of brightness and revealing the electron corona structures.

\subsection{Nonradial Height-Time Plots}
\label{sec:htt-plots}

\begin{figure}[h]
        \centering
        \begin{interactive}{animation}{Figure1_movie.mp4}
         \includegraphics[width=0.85\linewidth]{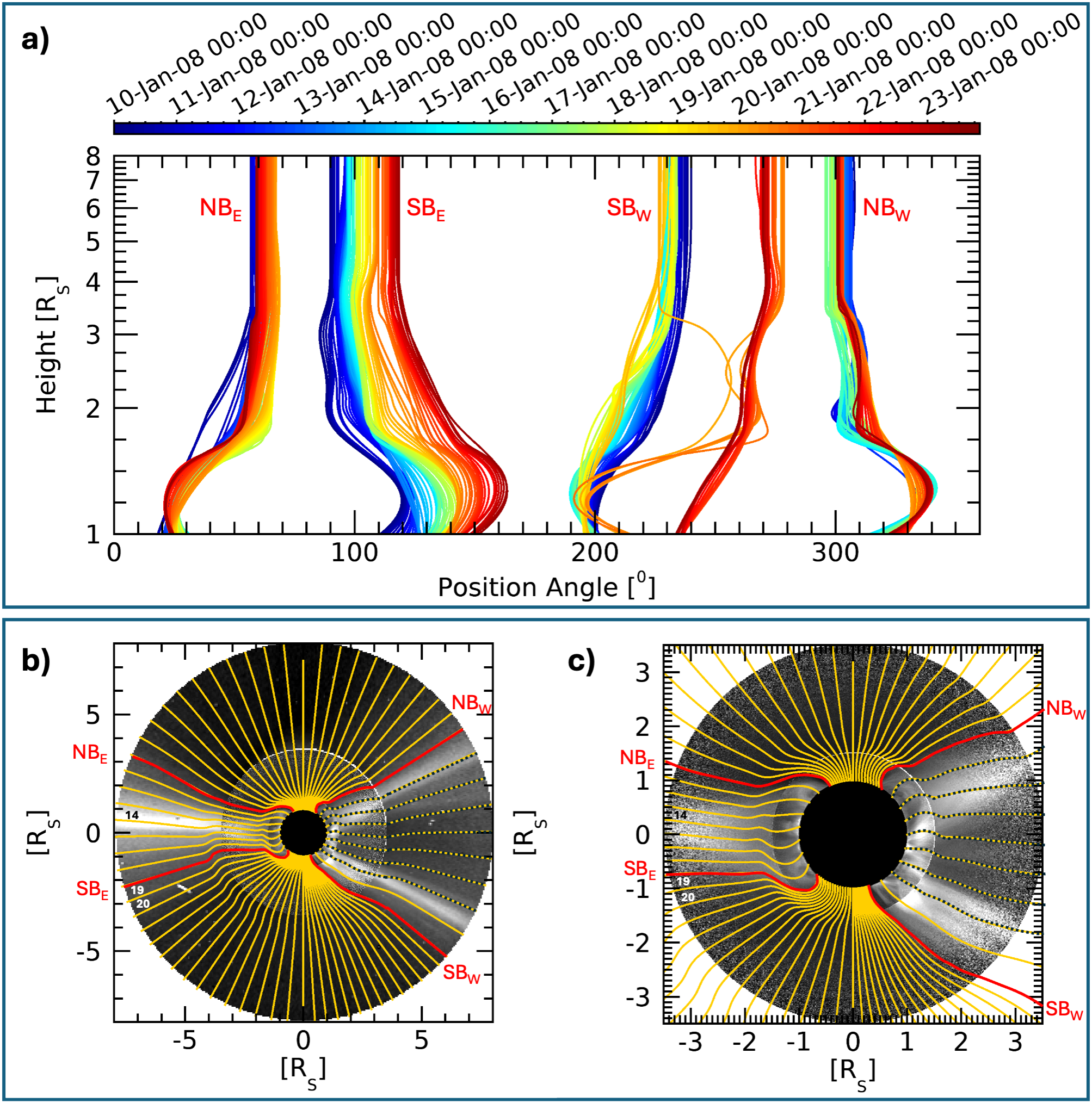}
        \end{interactive}
        \caption{Time evolution of north/south nonradial East/West limb streamer profiles ($NB_E/SB_E$ and $NB_W/SB_W$) detected through the STEREO/SECCHI suite between 1.0 and 8.0 $R_\odot$ in polar coordinates (panel a). The same profiles are over-imposed on the EUVI–COR1–COR2 composites processed with the NRGF to reveal the streamer profiles (panel b). Panel c, same as panel b but with a closer view of EUVI and COR1 observations. The yellow-black dotted lines show an example of paths excluded from the analysis during certain time intervals. A movie of this figure is available online. The animation spans 2008 January 10–23 at a rate of 4 hours per frame (3 s total duration).}
         \label{fig:1_nonradial_profiles}
    \end{figure}

Using the method presented in \citet{Alzate2023}, we generated nonradial profiles of propagating disturbances (PDs).  The method makes use of our advanced image processing techniques to identify streamer boundaries in solar images in polar coordinates, reinforced by the comparison with boundaries extracted from tomography brightness reconstruction.  From the identified boundaries we then generated nonradial Height-Time (Ht-T) plots along nonradial paths (see \citet{Alzate2023} for details).  For this study, we defined nonradial paths within the streamer as well as outside the streamer as seen in the POS. We used as reference points the solar North, solar South, and the identified boundaries of the streamers in the POS. Figure \ref{fig:1_nonradial_profiles}a shows, for the period under investigation, the time evolution of the north/south boundary of the streamer visible in the East ($NB_E/SB_E$) and West ($NB_W/SB_W$) limb. Then, we divided the regions in between the reference directions in a number of paths such that the average path width was 5$^\circ$ at 8 $R_{\odot}$. This choice led to a total of 67 paths (Figure \ref{fig:1_nonradial_profiles}b–c) and led to a good comparison with previous results in radial Ht-T plots in the COR1 and COR2 FOV. We count the path starting from 0 at the solar North and in increasing number anti-clockwise (the number of paths discussed in this manuscript are marked in panels b - c).  For the connection between EUVI and COR1 observations, we created nonradial paths with an average width of 5$^\circ$ at 1 $R_{\odot}$. This choice led to a total of 71 paths. Note that the streamer on the left (East) appears to expand towards the end of the time series, and the one on the right (West) splits at the beginning of the time series due to the presence of a pseudo-streamer \citep[see tomography reconstruction in][]{Alzate2023}. Issues arising from this effect can clearly be seen in the online movie associated with Figure \ref{fig:1_nonradial_profiles} when nonradial paths are not well aligned with the streamers. The yellow-black dotted lines in panels b - c show examples of excluded paths.  Results from these nonradial paths during these periods are excluded from our analysis. Note that, toward the end of the time interval, the nonradial procedure is properly identifying the northern streamer in the West limb, so results from those paths are included in our analysis. Following the improvements made to our methodology for tracking PDs and calculating relative velocity and acceleration profiles, we effectively defined nonradial paths for the entire FOV of the SECCHI suite (comprising EUVI, COR1, and COR2). Figure \ref{fig:2_nrhtt_plot_and composite}a shows an example of such nonradial Ht-T plots comprising EUVI, COR1, and COR2 FOV for the time period from 19 to 24 January and path 20 corresponding to approximately 91$^{\circ}$–164$^{\circ}$ in position angle. A bright track, corresponding to one of the fast PDs, crossed the FOV of the entire SECCHI suite and is marked here with arrows. Equivalent arrows are used to point to the same PD in panels b - d as observed in processed EUVI (BFF + NRGF), COR1 (BFF) and COR2 (BFF) images, as well as in panel e - g showing the same composites in polar coordinates. A movie for the entire analysis period is available online.

\subsection{PDs Analysis}
\label{sec:outflow}

\begin{figure}[h]
        \begin{interactive}{animation}{Figure2_movie.mp4}
        \includegraphics[width=0.9\linewidth]{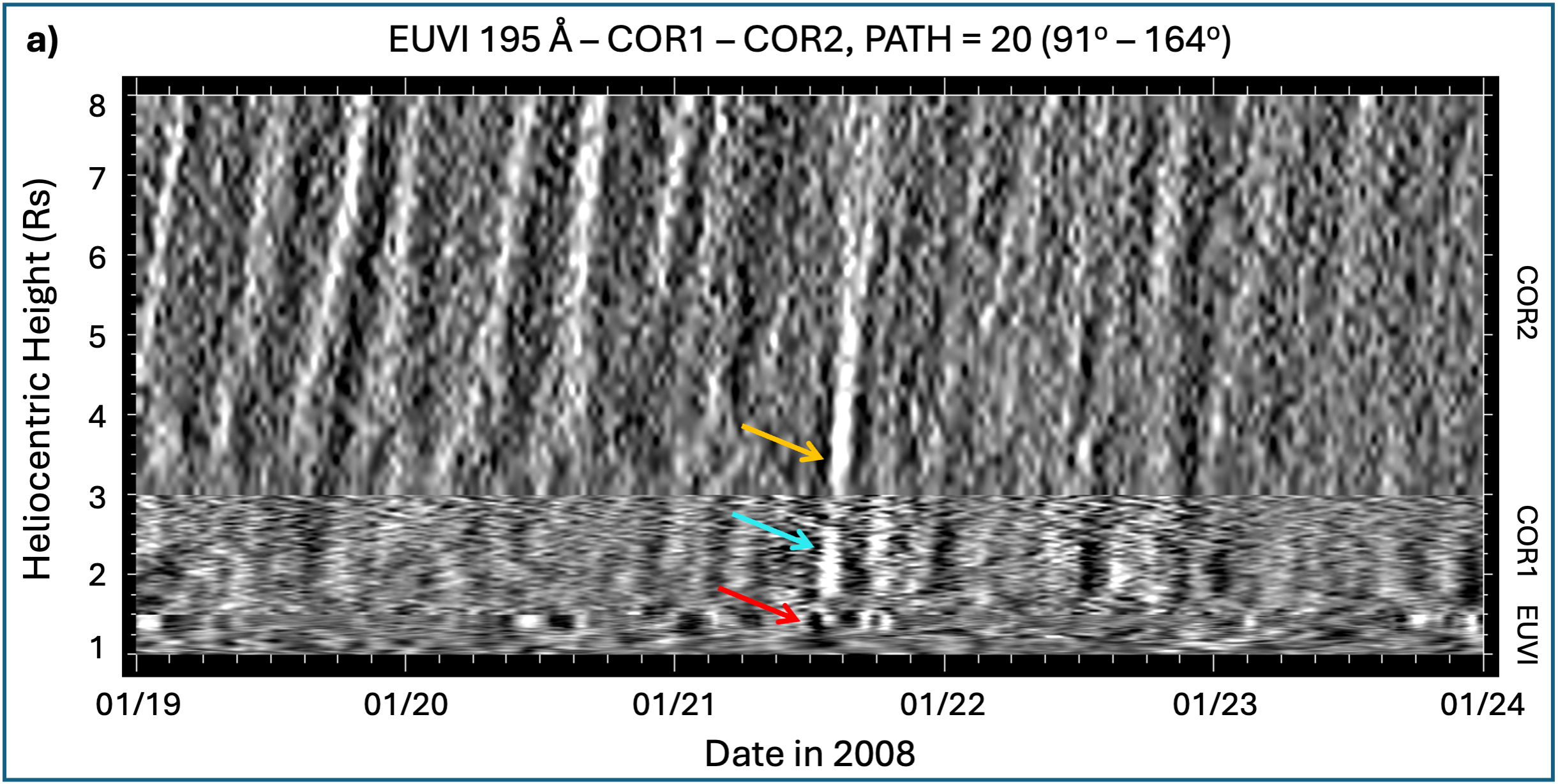}

        \vspace{0.25cm}
        
        \includegraphics[width=0.9\linewidth]{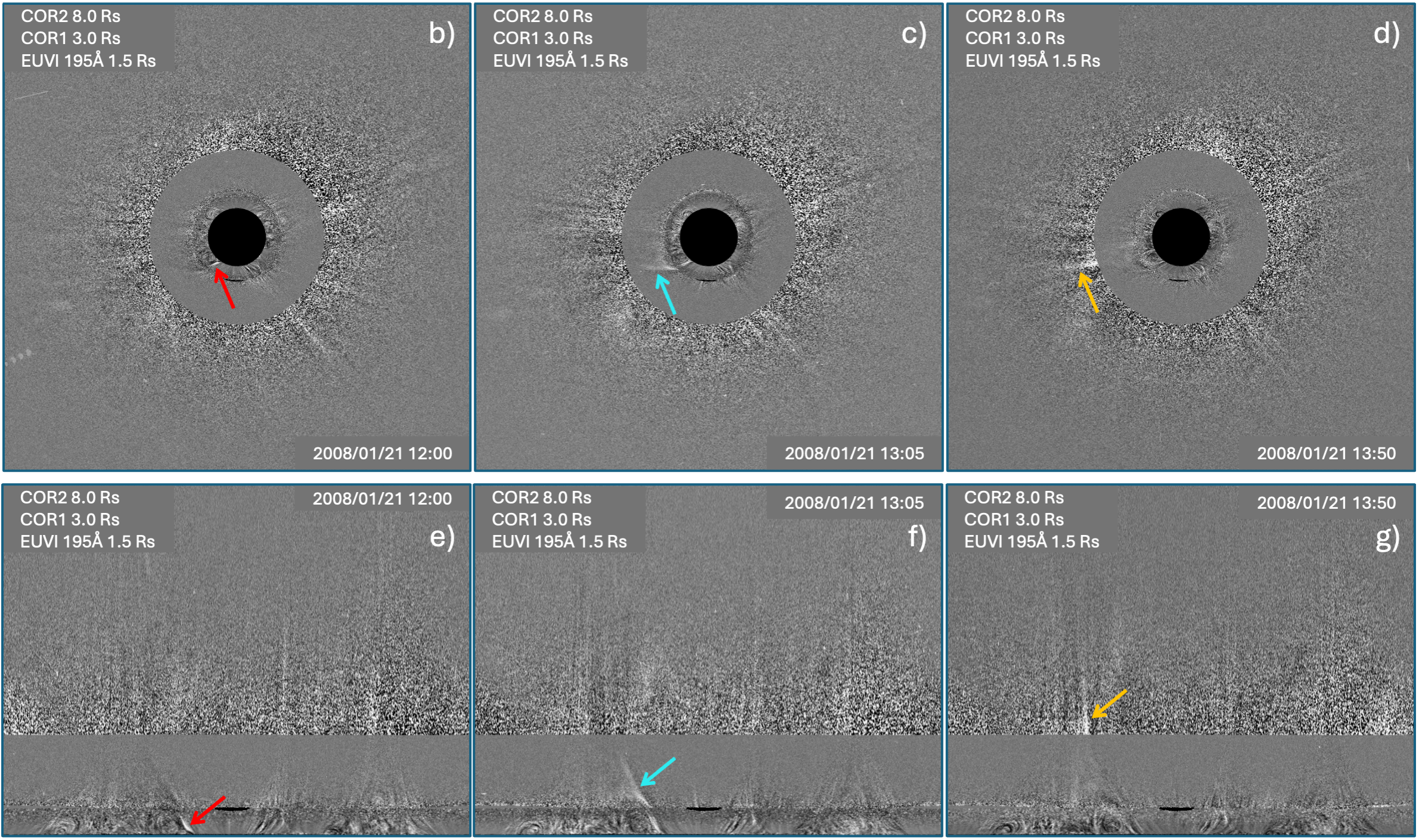}
        \end{interactive}
        \caption{An example of one of the faster PDs in a nonradial Ht-T EUVI-COR1-COR2 plot composite with corresponding PDs in EUVI-COR1-COR2 composite images.  The colored arrows in the Ht-T plots correspond to like-colored arrows in the images.  A movie of the image composites spanning the entire period of analysis is available online. The animation spans 2008 January 10–23 at a rate of one frame per hour (13 s total duration).  A version of the movie at higher cadence (5 minutes per frame; 323 s total duration) is available at \url{https://zenodo.org/uploads/11211569}\;\citep{Alzate2024}.}
        \label{fig:2_nrhtt_plot_and composite}
    \end{figure}

\begin{figure}[h]
        \centering
         \includegraphics[width=\linewidth]{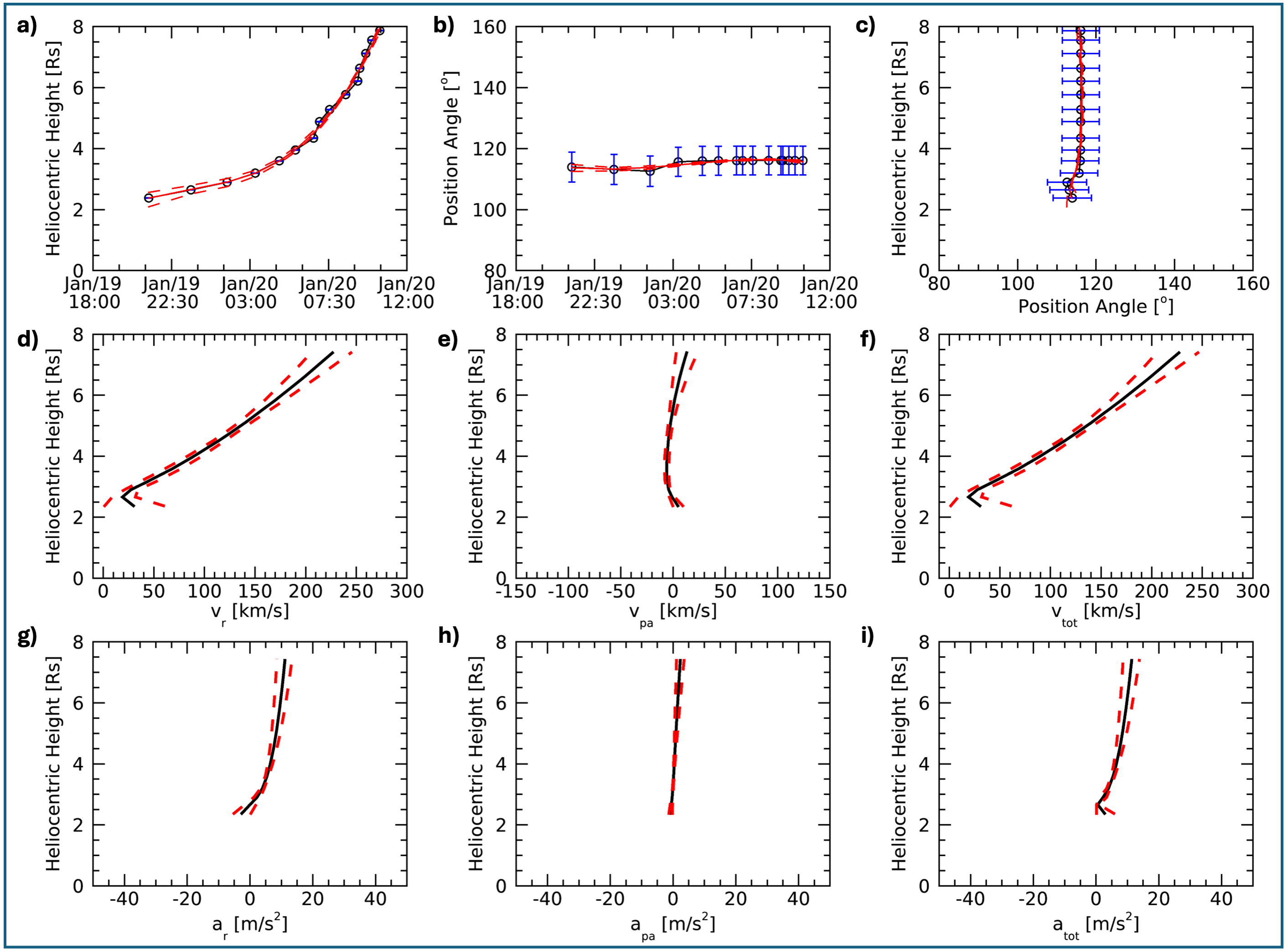}
        \caption{Output of the tracking method for one of the PDs shown in Figure \ref{fig:2_nrhtt_plot_and composite}. Top row, left to right: a) height versus time, b) position angle versus time, c) height versus position angle obtained from the point and click analysis with relative error bars. Red lines are the best fit resulting from the bootstrapping method. Middle row, left to right: d) height profiles of radial velocity, e) position angle velocity, and f) POS speed bounded by the absolute error (dashed lines). Bottom row, left to right: g-i) same as second row but for acceleration profiles.}
         \label{fig:3_tracking_example}
    \end{figure}

We improved a previously developed methodology aimed at tracking PDs in Ht-T plots and estimating relative velocity and acceleration profiles \citep{Byrne2013}.  The code extracts times and heights at specific pixels through a point-and-click approach.  Then a bootstrap approach is applied to fit a third-order polynomial function to the data and extract a smoothed Ht–T profile for each transient and the relative velocity and acceleration profile.  The observed tracks, especially in the low corona, manifested slope changes that were not well represented by the more often used paraboloid functions.  The new methodology can also extract the corresponding position angle (not constant because of the nonradial path) and effectively evaluate the projected velocity and acceleration in the POS.  Figure \ref{fig:3_tracking_example} shows an example of the results of the methodology tracking one of the PDs shown in Figure \ref{fig:2_nrhtt_plot_and composite}. The errors associated with the height and position angle are the ones resulting from the binning choice in the construction of nonradial Ht-T plots.  These errors are further propagated to estimate the velocity and acceleration profiles.

\subsection{Spectral Analysis}
\label{sec:spectral}

We performed a spectral analysis on time intervals characterized by numerous clustered PDs to check for occurrence of periodic release of plasma using the procedure by \citet{DiMatteo2021}.  Briefly, after zero padding the time series to reach two times the original length, we estimated the power spectral density (PSD) via the adaptive multitaper method \citep[MTM;][]{Thomson1982} with time-half bandwidth product NW = 3 and number of tapers K = 5.  Then, via a maximum likelihood criterion, we estimated the continuous PSD background fitting a pan-spectrum \cite[PNS,][]{Liu2020} function to the original PSD \cite[raw+PNS combination, see also][]{DiMatteo2021}.  The ratio between the PSD and the estimated background constitutes the $\gamma$ values. The MTM provides an additional independent statistical test, the F-test, to check for the presence of phase coherent periodic fluctuations in a time series. We impose an 80\% confidence threshold to identify the frequency of significant PSD and F-test enhancements, signature of periodic fluctuations in the time series ($\gamma$+F test). The combination of the two tests significantly reduces the number of false positives, typically by a factor of two \citep{DiMatteo2017,DiMatteo2021}, so it can be assumed that the effective confidence level is close to 90\%.

\section{Results/Observations} 
\label{sec:results}

Our analysis of PDs led to the identification of 1132 tracks. One of the main difficulties of our analysis was the minimization of possible artifacts arising from the nonradial plasma motion in the low corona and possible line-of-sight effects. The nonradial procedure has already been proven to minimize these issues \citep{Alzate2023}.  In this work, however, we further clean the data by imposing additional constraints. First, we collected tracks that occurred at similar height ranges, position angles, and times. Nonradial motion of plasma beyond the defined nonradial path and/or size scales of PDs occupying more than one path could lead to multiple tracks. We addressed these complications by comparing each point of one track against each point of neighboring tracks and collecting tracks for which at least one pair of points occurred with a time difference of less than 30 minutes, a difference between logarithmic heights less than 0.05, and position angle differences of less than 10$^\circ$. Furthermore, for PDs moving away from the sun, we only considered the ones whose tracks reach heights beyond 2.5 $R_{\odot}$. These selection criteria reduced our dataset to 417 unique outward PDs (OPDs; away from the sun) and 31 unique inward PDs (IPDs; toward the sun). In the following sections, we describe their properties.

\subsection{Velocity and Acceleration Profiles}
\label{sec:outflow}
\begin{figure}[h]
        \centering
         \includegraphics[width=0.8\linewidth]{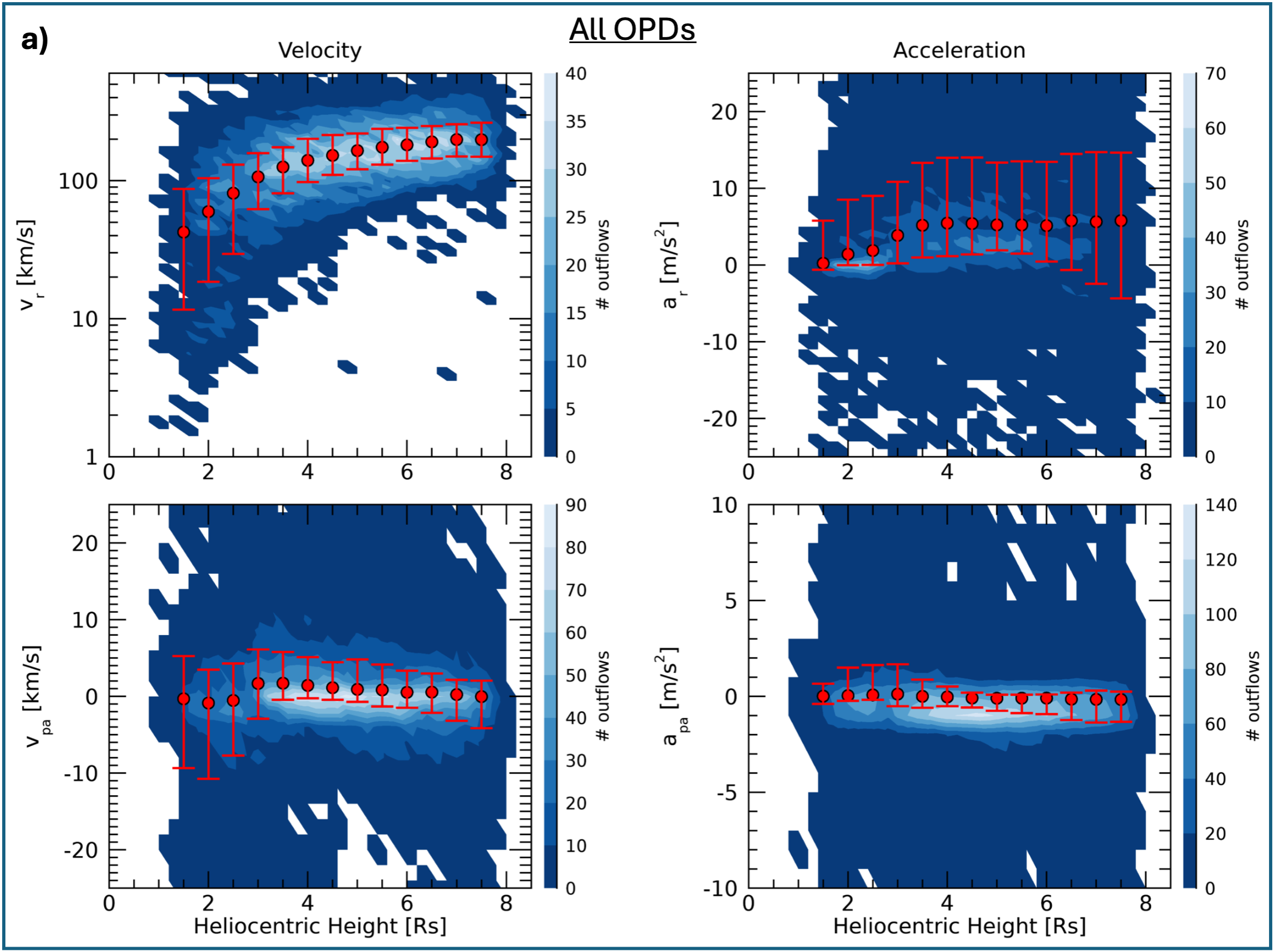}

         
         \includegraphics[width=0.8\linewidth]{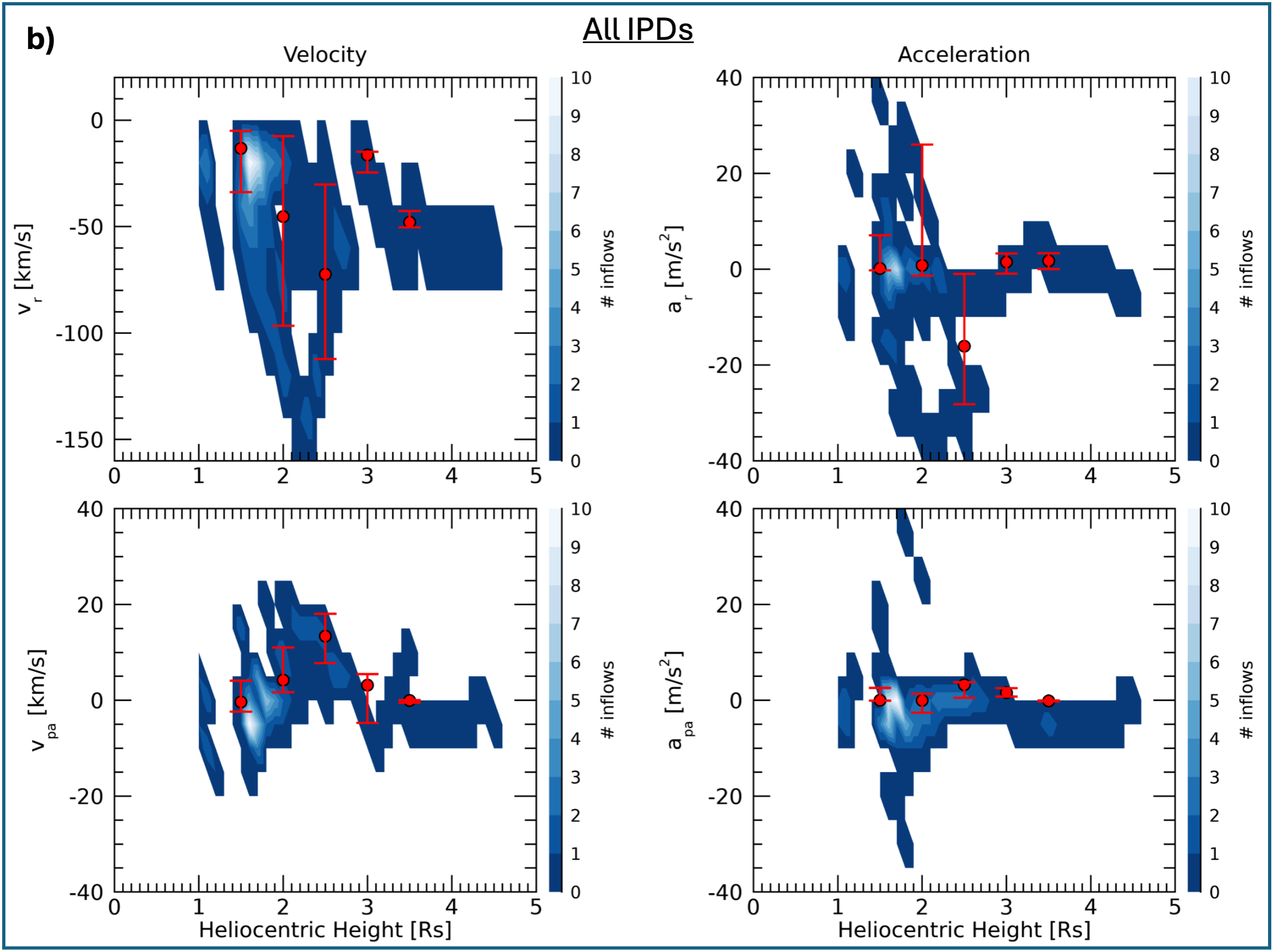}
        \caption{Velocity and acceleration profiles for a) all OPDs and b) all IPDs. In both panels, top row shows the profile of $v_r$ and $a_r$, while the bottom row shows the $v_{pa}$ and $a_{pa}$ profiles. The color scale indicates the number of OPDs while the errorbars indicate the median value (red dot) and interquartile range (red bars) at height bins of 0.5 $R_\odot$.}
         \label{fig:4_velacc_out_in}
    \end{figure}

We first investigated the height profiles of speed and acceleration along the radial ($v_{r}$ and $a_{r}$) and position angle ($v_{pa}$ and $a_{pa}$) direction for the identified OPDs and IPDs. For the position angle velocity and acceleration, we note that most of the OPDs get closer to the position angle of the streamer center as they move outward. Therefore, we separated OPDs based on the limb where they were observed and set the sign of $v_{pa}$ so that positive values indicate propagation toward the center of the streamer observed in the respective limb; the same is true for $a_{pa}$. Figure \ref{fig:4_velacc_out_in} shows bidimensional distributions with color scales indicating the number of OPDs at a certain height with a certain speed or acceleration. The distribution trends are obtained by collecting the median value (red dot) and the interquartile range (red bar in the Figure and subscript/superscript in the following text) of the distribution in height bins of 0.5 $R_\odot$. OPDs have radial speeds of $42.4_{-30.8}^{+44.58} km/s$ at 1.5 $R_\odot$ and gradually accelerate to $198.0_{-48.9}^{+64.8} km/s$ at 7.5 $R_\odot$. The radial acceleration is $0.2_{-0.8}^{+5.6}m/s^2$ at 1.5 $R_\odot$ and gradually increases to $\approx$ 4.0 $R_\odot$, after which it remains stable assuming values of $5.8_{-10.2}^{+8.9}m/s^2$ at 7.5 $R_\odot$. The median $v_{pa}$ shows slight negative values below 2.5 $R_\odot$ after which it assumes positive values (motion toward the center of the streamer) starting with a speed of $1.7_{-4.6}^{+4.4}km/s$ at 3.0 $R_\odot$ and decreasing to almost null values at 7.5 $R_\odot$. The median $a_{pa}$ values showed accordingly slight positive and negative values respectively below and above 3.5 $R_\odot$. IPDs are concentrated mainly below 2.0 $R_\odot$ with radial speed and acceleration of $-13.3_{-20.5}^{+8.3}km/s$ and $0.1_{-0.4}^{+6.9}m/s^2$ at 1.5 $R_\odot$ and no significant $v_{pa}$ or $a_{pa}$. We show velocity and acceleration profiles of IPDs forming at greater heights, but their low number makes the median $v_{pa}$ and $a_{pa}$ values above 2.0 $R_\odot$ questionable.

\begin{figure}[h]
        \centering
         \includegraphics[width=0.8\linewidth]{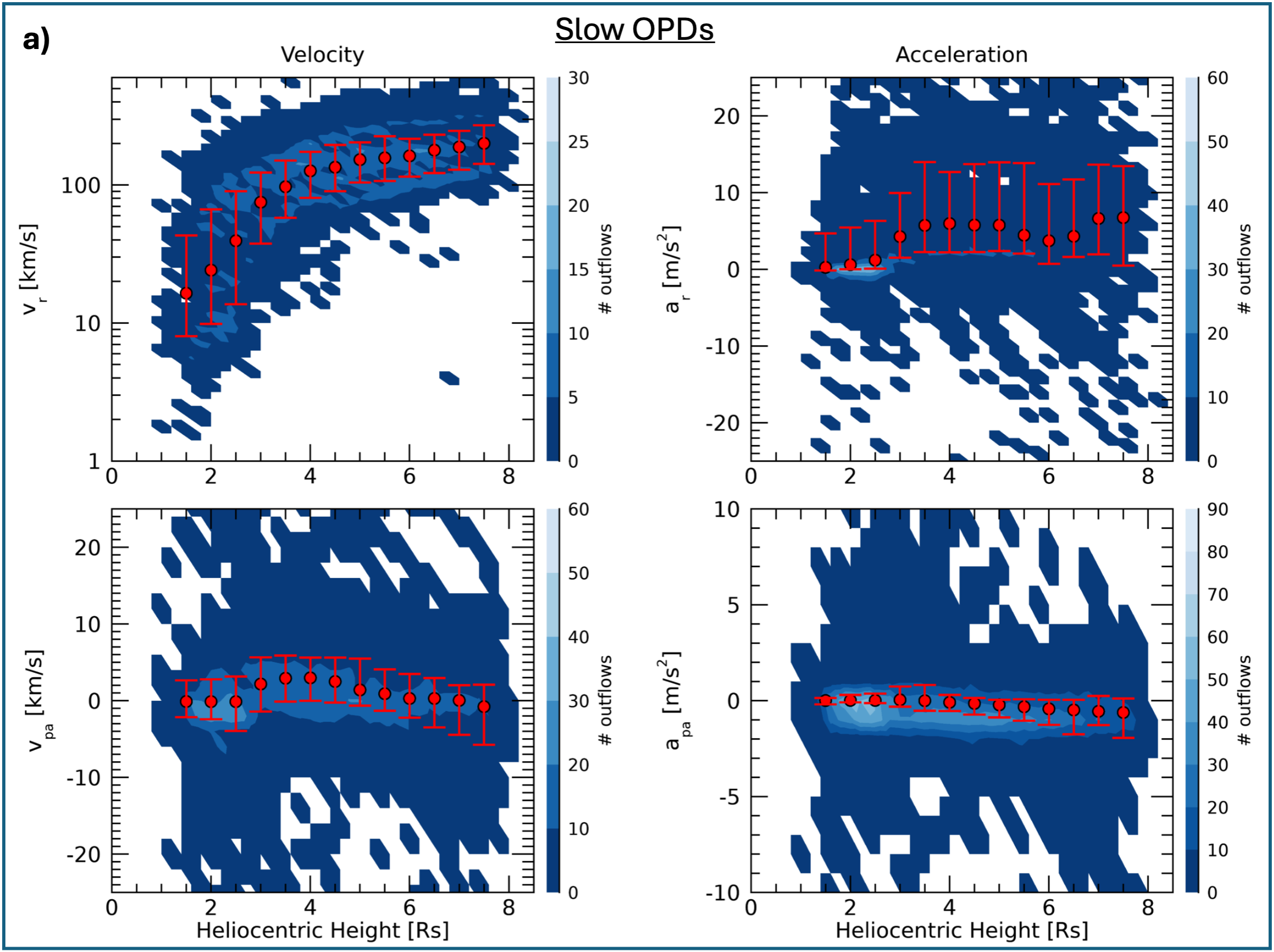}

         
         \includegraphics[width=0.8\linewidth]{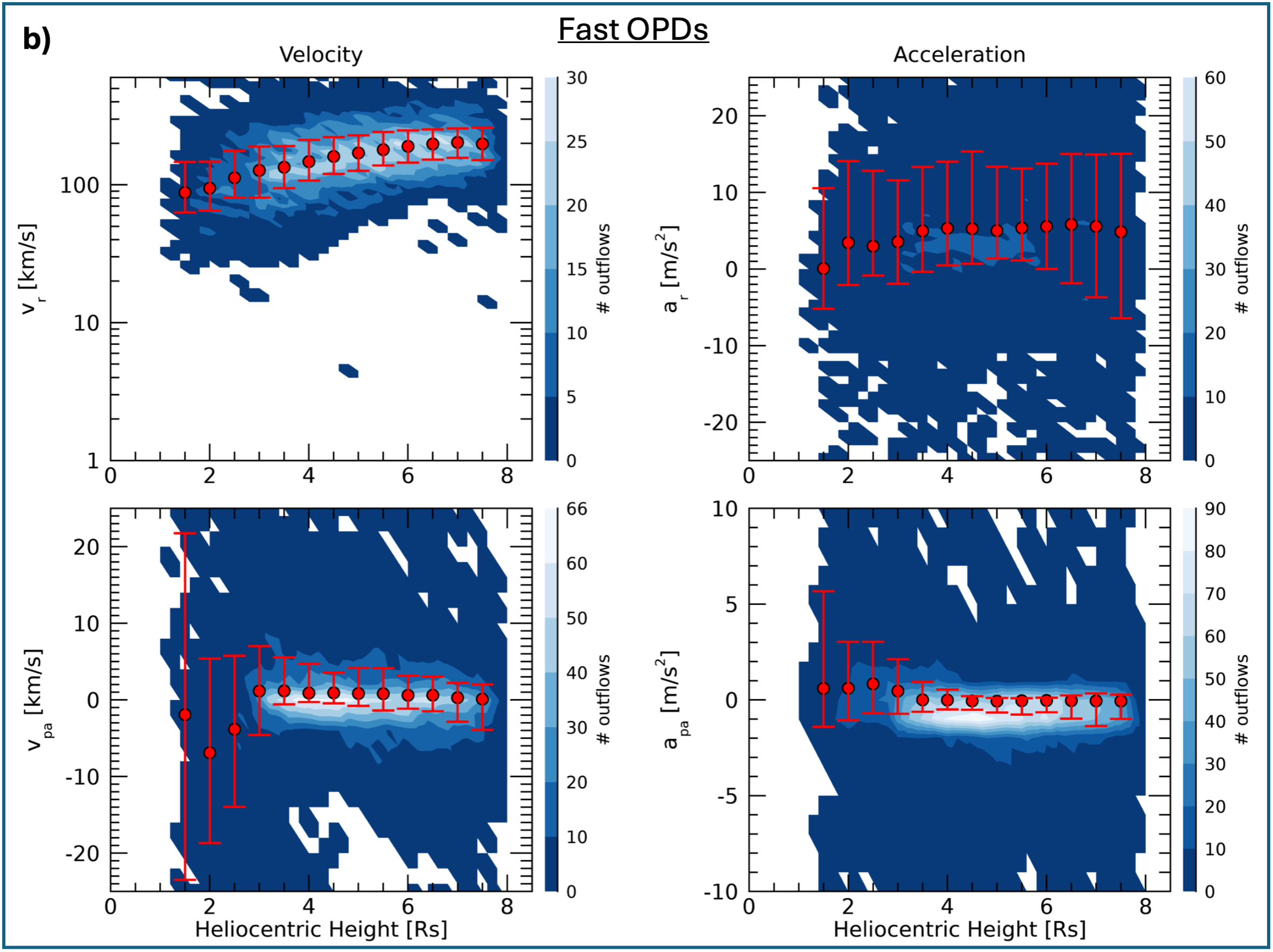}
        \caption{Velocity and acceleration profiles for a) slow OPDs and b) fast OPDs. In both panels, top row shows the profile of $v_r$ and $a_r$, while the bottom row shows the $v_{pa}$ and $a_{pa}$ profile. The color scale indicates the number of OPDs while the errorbars indicate the median value (red dot) and interquartile range (red bars) at height bins of 0.5 $R_\odot$.}
         \label{fig:5_velacc_slow_fast}
    \end{figure}

The radial velocity of both OPDs and IPDs (although less clearly) manifests itself in two peaks of main occurrence around 2.0 $R_\odot$. For OPDs, we obtain a higher occurrence at $\approx$10--20 $km/s$ and $\approx$50 $km/s$, while for IPDs there is a clear peak at $\approx$-20 $km/s$ and some indication of another group at $\approx$-120 $km/s$. These suggest the presence of two categories of transients, slow OPDs and fast OPDs, which have already been discussed in previous works identifying transients in the low corona \citep{Alzate2021,Seaton2021}. To separate the two populations, we applied a cluster analysis \cite{Everitt1993} on the minimum height and radial speed of each unique outflow.  We did this using the function {\fontfamily{cmtt}\selectfont CLUSTER} (within the \textit{ Interactive Data Language} (IDL) based system) with two weights set at a height of 1.5 $R_\odot$ and radial velocity of $\approx$10 $km/s$ and $\approx$80 $km/s$.  Note that varying the starting weights has negligible effects on the following results. The speeds and accelerations are shown in Figure \ref{fig:5_velacc_slow_fast} in the same format as Figure \ref{fig:4_velacc_out_in} for 174 slow OPDs (panel a) and 243 fast OPDs (panel b). There are some differences from the collective properties of all the OPDs. The slow OPDs have a lower initial radial speed of $16.4_{-8.4}^{+26.6}km/s$ at 1.5 $R_\odot$ and accelerate up to $200.1_{-57.9}^{+71.1}km/s$ at 7.5 $R_\odot$. The radial acceleration profile shows a steep increase around $\approx$3.0 $R_\odot$ going from almost null values to $6.0_{-3.7}^{+6.7}m/s^2$ at 4.0 $R_\odot$ after which the median $a_{r}$ shows values within $4.3$ and $6.7m/s^2$. As for the main population, the median $v_{pa}$ shows slight negative values below 2.5 $R_\odot$ after which it assumes positive values (motion toward the center of the streamer) peaking with a speed of $3.0_{-3.0}^{+2.6}km/s$ at 4.0 $R_\odot$ and decreasing to almost null values at 7.5 $R_\odot$. The median $a_{pa}$ values showed accordingly slight positive and negative values respectively below and above 4.0 $R_\odot$. The fast OPDs show significantly higher radial speeds starting from $87.8_{-24.8}^{+59.1}km/s$ in the low corona at 1.5 $R_\odot$ up to $197.8_{-46.7}^{+61.8}km/s$ at 7.5 $R_\odot$. Interestingly, the acceleration profile remains almost constant for heights greater than 2.0 $R_\odot$ with median values within $3.0$ and $5.8m/s^2$. The median $v_{pa}$/$a_{pa}$ assume negative/positive values below 2.5 $R_\odot$ but they are associated with large uncertainties. Above 2.5 $R_\odot$, we again observe a $v_{pa}$ distribution skewed toward positive values (motion toward the streamer center) and almost null $a_{pa}$. Finally, note that in the radial velocity of slow OPDs we observe an isolated small population with very low speeds (below $\approx$15 $km/s$) which shed some doubts about their interpretation as real OPDs. Additionally, the speed values for this population are just above the ones expected for a streamer motion toward the POS as the sun rotates. For completeness, in Appendix A, we show the kinematic results that we obtain after excluding this population from our analysis. Briefly, the velocity and acceleration profiles remained unaltered except for the median $v_r$ and $a_r$ profiles which showed higher values at lower heights, but with speeds still distinct from the fast OPDs.

The radial velocity ranges of the slow and fast OPDs include those observed in previous investigations.  \citet{Alzate2021} reported an average velocity of $\approx$ 4.4 $km/s$ for slow OPDs and an average of $\approx$ 131 $km/s$ for fast OPDs with radial tracking, while \citet{Alzate2023} reported an average velocity of $\approx$ 3.2-6.4 $km/s$ for slow OPDs and an average of $\approx$ 95-135 $km/s$ for fast OPDs with nonradial tracking.  \citet{Seaton2021} reported an average of a few tens of $km/s$ for slow OPDs and an average of $\approx$50-150 $km/s$ for fast OPDs.

\subsection{Formation Location}
\label{sec:outflow}
\begin{figure}[h]
        \centering
         \includegraphics[width=0.92\linewidth]{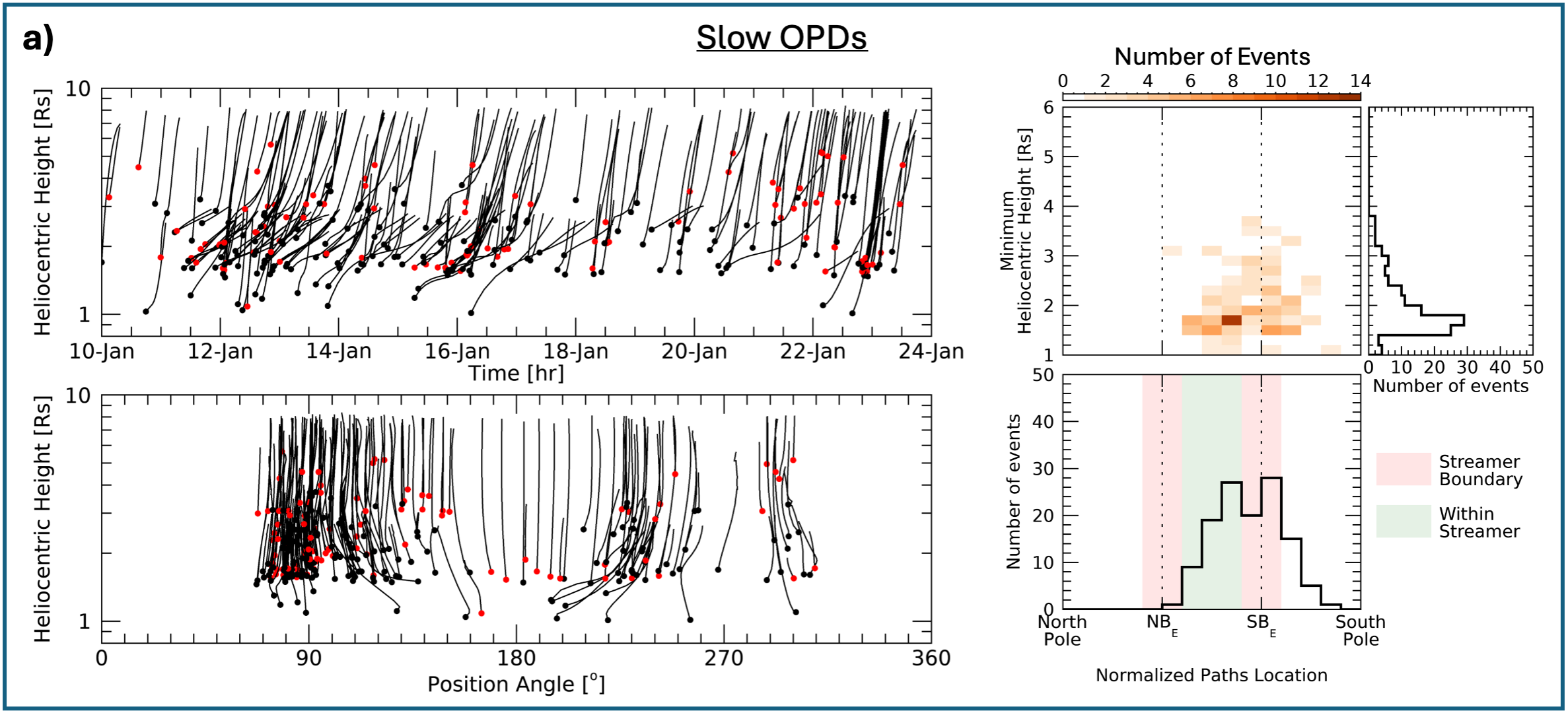}
         
         
         \includegraphics[width=0.92\linewidth]{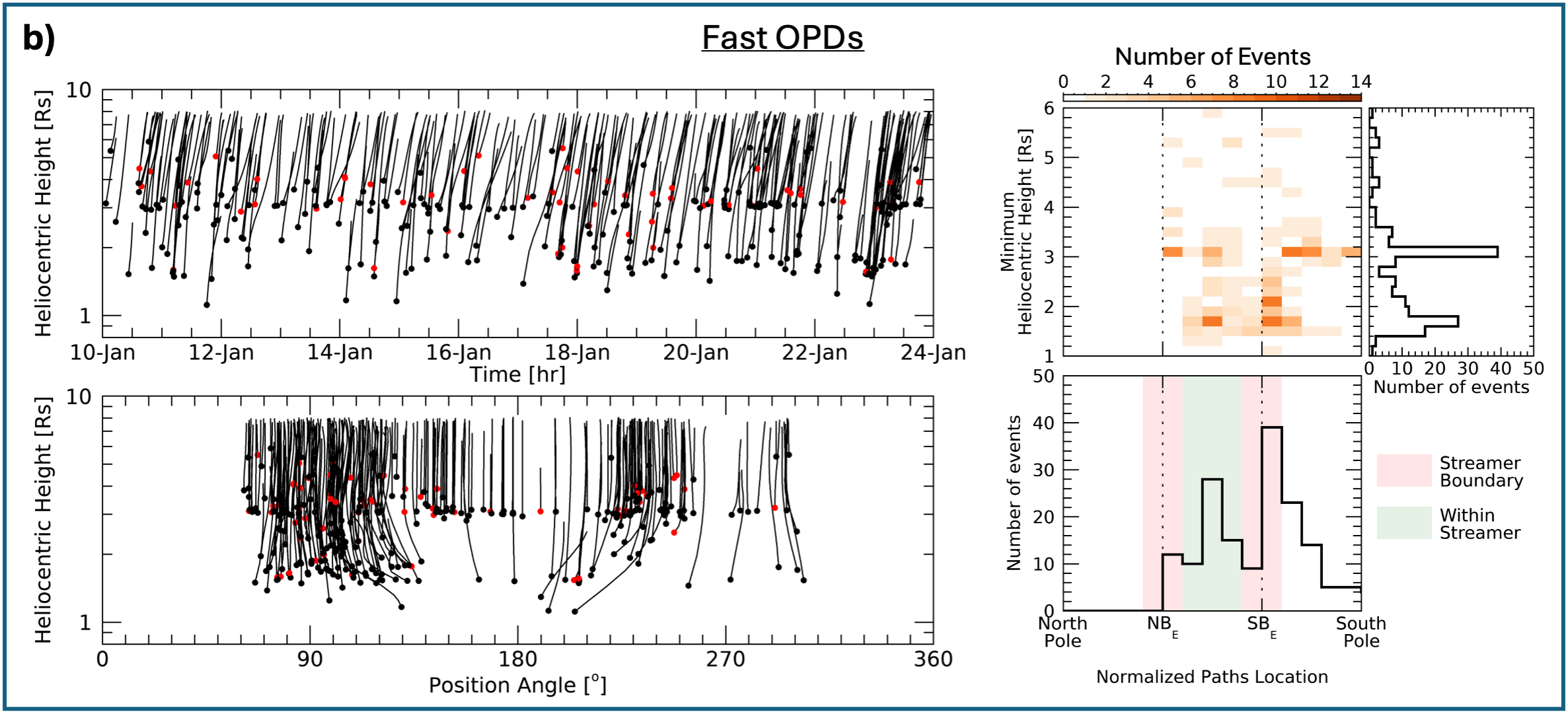}

         
         \includegraphics[width=0.92\linewidth]{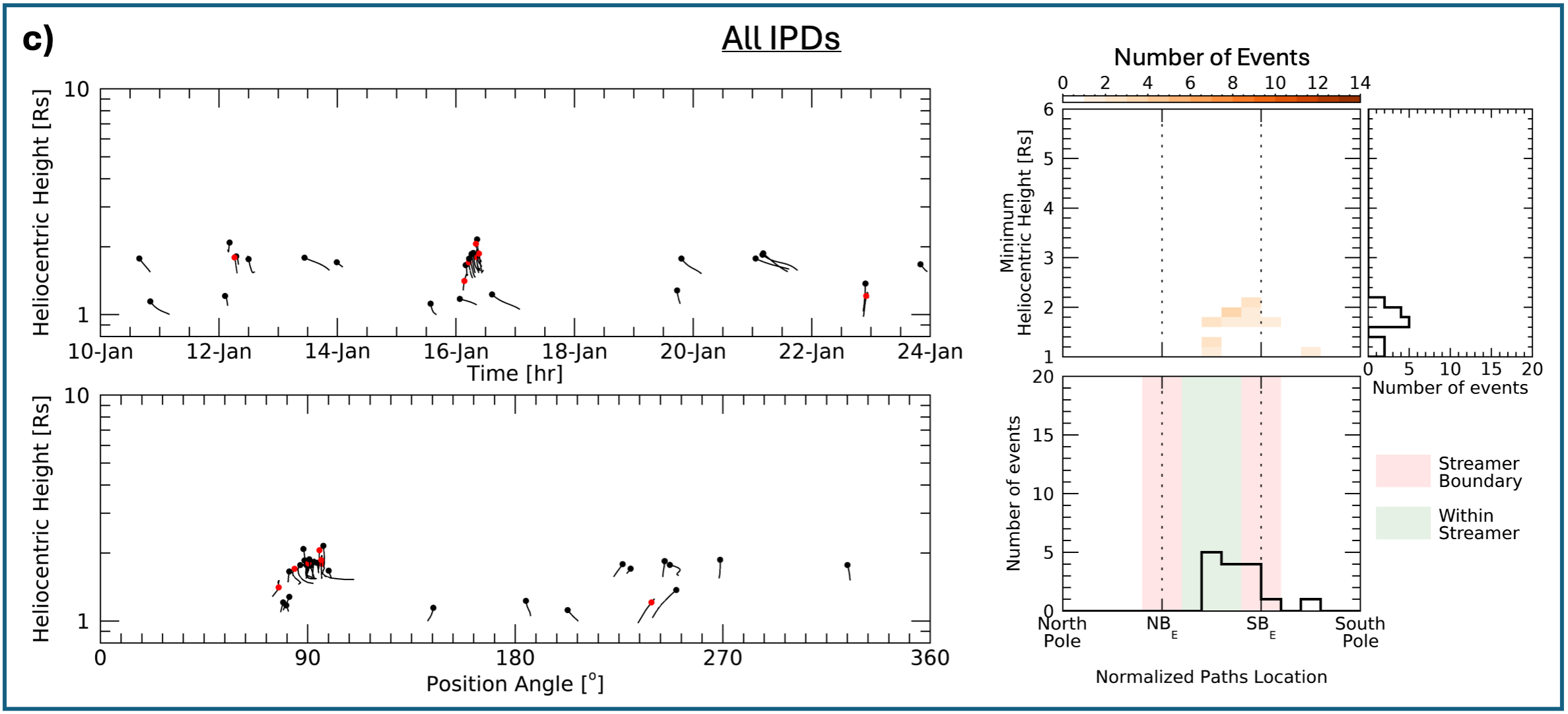}
        \caption{Spatio-temporal distribution of slow OPDs (panel a), fast OPDs (panel b), and IPDs (panel c) with corresponding 1D and 2D distribution of relative formation heights and location with respect to the East streamer boundaries (see also Figure \ref{fig:1_nonradial_profiles}). Black dots in the Ht-T and Ht-pa plots mark the starting point of unique OPDs and IPDs, while the ones being grouped into a main OPD/IPD are marked with red dots.}
         \label{fig:6_formation_height}
    \end{figure}

Next, we investigated the formation location of the OPDs, classified as slow and fast OPDs, and IPDs. For each category, we reported in Figure \ref{fig:6_formation_height} the tracks (black lines) we detected in the Ht-T plots and the corresponding location in polar coordinates. Black dots mark the starting point of a unique outflow as defined by our criteria. Groups of OPDs that are associated with a main unique outflow starting at lower heights are marked by red dots at their starting point. While we consider the starting point height as the formation height of the unique OPDs and IPDs, we use the path number in which the transients are observed to investigate the position angle distribution. The latter approach allows to qualitatively investigate the location of the transients with respect to the boundaries of the streamer. In fact, even though the streamer shape evolves over time, the number of paths that we define within it remain constant. We perform a detailed analysis only for the streamer in the East limb for which we have continuous coverage for the entire period in analysis. We normalize the path number in the areas north, south, and within the East streamer boundaries $NB_E$ and $SB_E$ by the respective number of paths. The paths adjacent to each identified streamer profile are defined as the boundary region. The resulting 2D distributions of the formation location are shown in Figure \ref{fig:6_formation_height} in color scale for each category, while the histogram shows the 1D distribution of the transients' formation height and path. Results for slow and fast OPDs manifest some significant differences. Slow OPDs form between 1.0 and 3.8 $R_\odot$, preferentially at $\approx$1.6 $R_\odot$. These OPDs appear more concentrated around the south streamer boundary ($SB_E$). Fast OPDs can originate up to 5.8 $R_\odot$ but preferentially at $\approx$1.6 $R_\odot$ and $\approx$3.0 $R_\odot$. Note that the dip in occurrence around 2.5 $R_\odot$ might be affected by instrumental effect since it corresponds to the inner edge of COR2 and it is where the COR1 data start to get noisy.  While the fast OPDs also lay around the south boundary, there is a higher occurrence rate within the streamer and around the north boundary with respect to the slow OPDs. A significant number of OPDs is observed around the south pole far from the streamer. A detailed investigation revealed that all the OPDs occurred around 23 January, right after the passage of a CME suggesting a likely connection of the fast OPDs in this region to post-eruption blobs \cite[e. g.,][]{Hess2017}.  For the IPDs, they appear within the streamer and around the south boundary while the histogram of formation height shows a peak at $\approx$1.8 $R_\odot$. The IPDs we identified are more sporadic but most of them appear to concentrate around 16 January between position angles 80$^\circ$ and 100$^\circ$. Interestingly, their occurrence follows the transit of a CME; the properties of the observed IPDs are consistent with the ones reported by \cite{Hess2017} for IPDs related to reconnection of a trailing current sheet. These IPDs were observed below heights of 2.0 $R_\odot$ with speeds ranging between 10 and 150 $km/s$ up to several days after the CME.

The presence of two preferential formation heights for OPDs could be an indication of two different kinds of release mechanisms of coronal plasma. OPDs originating at $\approx$3.0 $R_\odot$ and more aligned to the center of the streamer could be more closely related to processes at the tip of streamers \cite[e. g., tearing mode;][]{Reville2020}. Interestingly, some OPDs originating at low heights find correspondence with the formation of IPDs both covering the range $\approx$1.6–1.8 $R_\odot$. When observed, OPD and IPD pairs suggest the occurrence of magnetic reconnection and/or interchange reconnection phenomena for the plasma release \cite[e. g.,][]{Sanchez2017b}. Note also that detailed investigations of inflows in remote sensing observations reported that they are mainly seen along bends of the coronal streamer belt \citep{Sheeley2001} suggesting that the inflows are easier to detect when the streamer belt is seen face-on \citep{Sheeley2001,Sanchez2017a}. The streamer configuration for the period under investigation shows only small bends. This can be seen from rotational tomography coronal density reconstruction which was reported for the same period in \citet{Alzate2023}. This is in agreement with our observed unbalance of detection between OPDs and IPDs. However, a more in depth analysis over different time periods is necessary to fully quantify the amount of OPD and IPD pairs and their relation to a magnetic reconnection scenario.

\subsection{Spectral Analysis}
\begin{figure}
        \centering
         \includegraphics[width=0.55\linewidth]{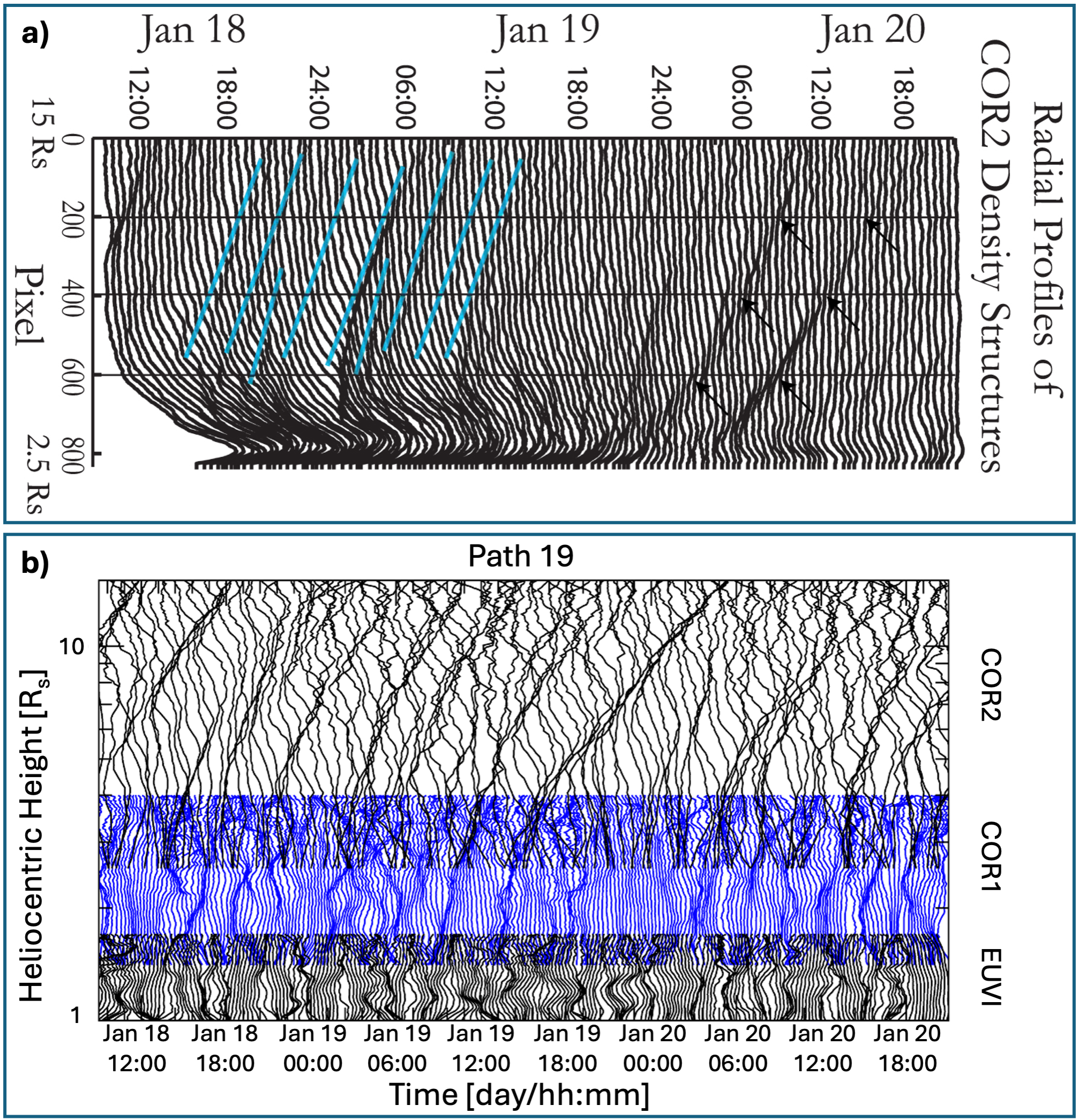}
        \caption{Panel a) Ht-T radial profiles for COR2 adapted from \citet{Viall2015}. Panel b) Nonradial Ht-T profiles for EUVI–COR1–COR2, radially normalized and centered in time (see text for details), for the same time period and location (path 19; see Figure \ref{fig:1_nonradial_profiles} for its location).}
         \label{fig:7_comparison_with_VV15}
    \end{figure}

The period we analyzed here had already been discussed in detail by \cite{Viall2015} in terms of periodic release of coronal plasma. Typical periodicities ranging between 65 and 100 min were identified in correspondence with the visible streamer in the East limb of the COR2 FOV. Given our ability to identify transients in the uninterrupted FOV using EUVI-COR1-COR2 observations, we extended the investigation on the occurrence of periodic plasma release down to the low corona. First, we compared the results of our methodology with the example time interval already discussed by \cite{Viall2015} in which 90 min periodicities were detected (panel a in Figure \ref{fig:7_comparison_with_VV15}). In order to make a direct comparison, we extracted brightness height profiles from the EUVI–COR1–COR2 observations to which we had applied the BFF procedure. To remove the radial brightness trend, we standardized the observations at each height for the period of interest, which is a simplified version of the NRGF technique. A simple smoothing over a sliding window of 13 points in height was applied to best reproduce the example of \cite{Viall2015}. To align the profile in time, we removed the average brightness of each profile and added the actual time of the observations. The resulting profiles for EUVI–COR1–COR2 observations are shown in the panel b of Figure \ref{fig:7_comparison_with_VV15} along a nonradial path that more closely overlaps with the radial path used by \cite{Viall2015}. Even though we used a different approach, we can recognize the denser number of OPDs at the beginning of the time interval. Most of the tracks visible in the COR2 FOV can easily be connected to tracks in the COR1, and subsequently the EUVI, FOV. Most of these transients originate at heights below $\approx$2.0 $R_\odot$.

\begin{figure}
        \centering
         \includegraphics[width=0.49\linewidth]{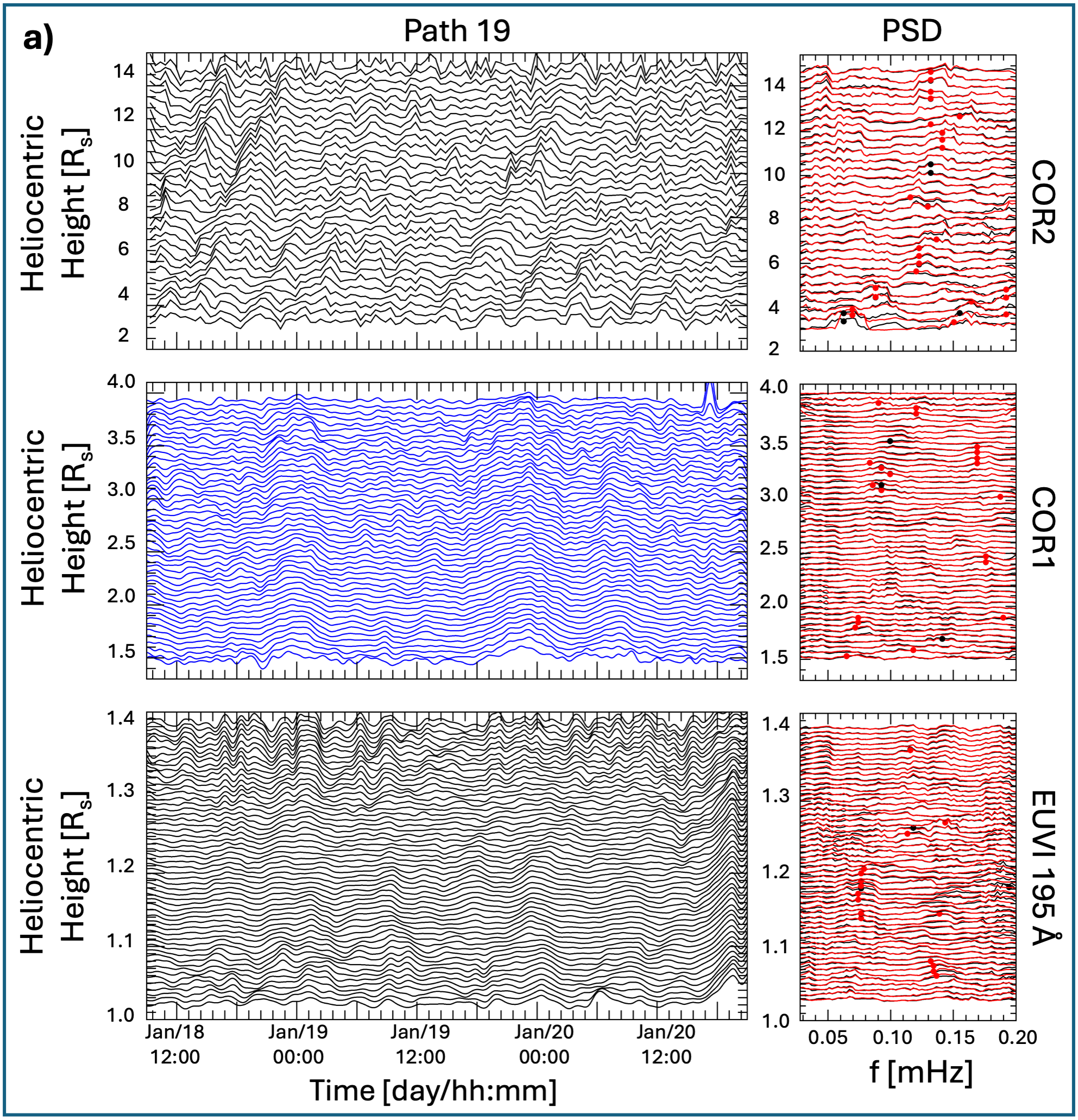}
         \includegraphics[width=0.49\linewidth]{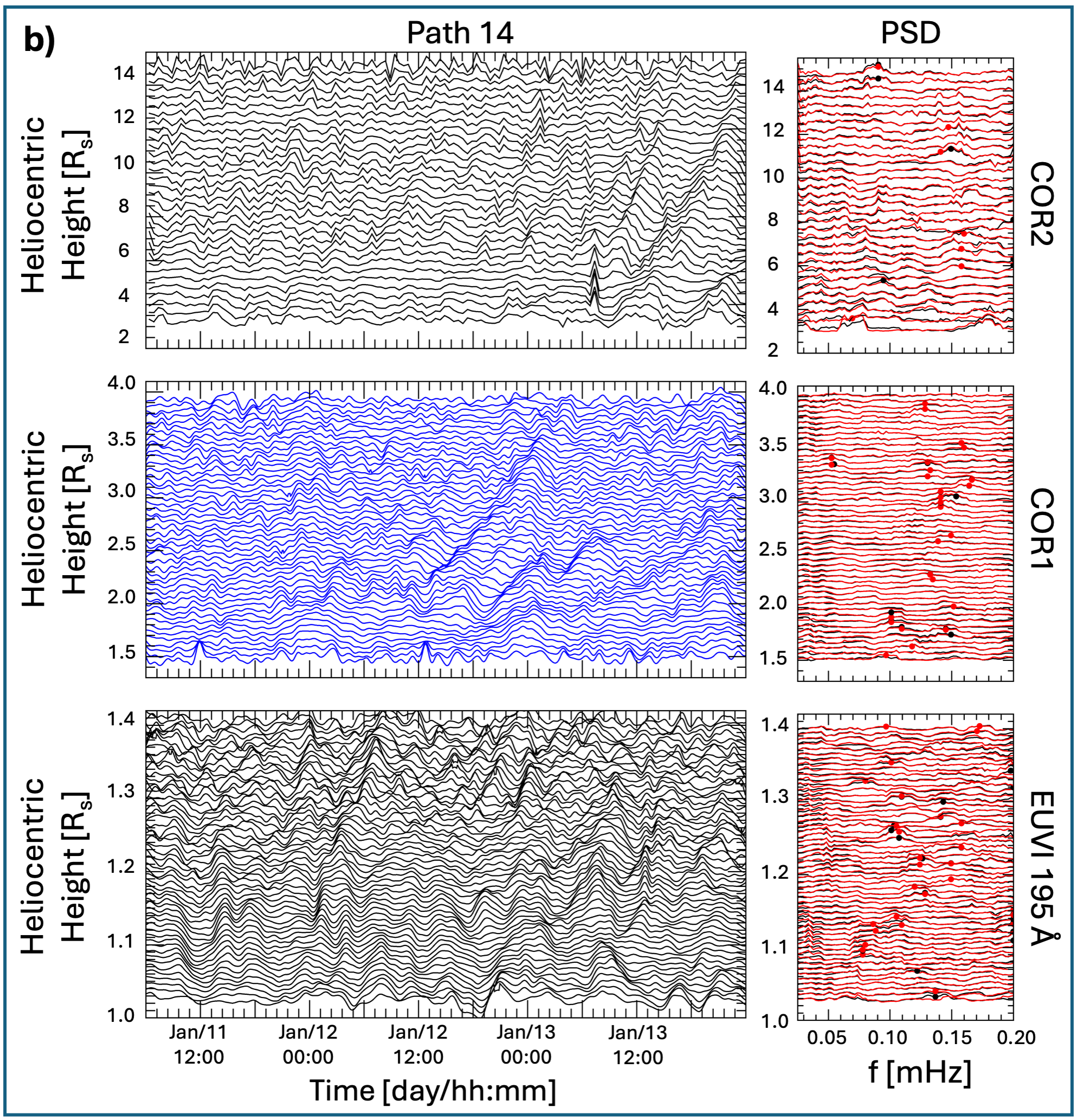}
        \caption{Panel a) On the left, brightness profiles at fixed heights in the EUVI, COR1,and COR2 FOV for the same time interval and path shown in Figure \ref{fig:7_comparison_with_VV15}. On the right, normalized PSD from original data (black) and BFF processed data (red) with dots marking the portion satisfying the $\gamma$+F test at the 80\% confidence level. Panel b) Same as panel a) but for the nonradial path 14 on January 11–14. See Figure \ref{fig:1_nonradial_profiles} for paths location.}
         \label{fig:8_spectral_analysis}
    \end{figure}

To perform the spectral analysis, we extracted brightness profiles at fixed heights from the BFF processed nonradial Ht-T plots. This is reported in Figure \ref{fig:8_spectral_analysis}a for the same time period discussed in Figure \ref{fig:7_comparison_with_VV15}. In this representation the OPDs are more clearly seen in the COR2 FOV but are fainter in COR1 and EUVI observations. The panels on the right in Figure \ref{fig:8_spectral_analysis}a show the power spectral density normalized to the identified background (see Section \ref{sec:spectral}) with dots identifying the significant portion of the PSD associated with brightness periodic fluctuations. To ensure that the BFF technique did not include any artifacts in our results, we compared the spectral analysis results from the BFF filtered data (red) with the one obtained from the original observations (black) on which we only applied the radial normalization and smoothing in height. In the COR2 FOV we identified periodicities in the 0.12–0.16 mHz (106–139 min) range at heights between 5.5 and 15 $R_\odot$ confirming the occurrence of the periodicity reported by \cite{Viall2015}. The PSD peak at $\approx$0.07 mHz close to the inner edge of COR2 is a known instrumental signal which shows up most strongly near the edges of the COR2 FOV \citep{Viall2015}. Moving to lower heights, we identified periodicities at progressively longer periods, that is 0.08–0.10 mHz (167–208 min) between 3.0 and 5.0 $R_\odot$ and 0.07–0.08 mHz (208–238 min) between 1.15 and 1.2 $R_\odot$. Similar patterns are observed in other time intervals and regions. Figure \ref{fig:8_spectral_analysis}b shows a similar spectral analysis for path 14 on January 11–14 during which OPDs were clearly visible in the EUVI–COR1–COR2 FOV (Figure \ref{fig:8_spectral_analysis}a). Spectral analysis results in Figure \ref{fig:8_spectral_analysis}b reveal periodicities in the 0.13–0.17 mHz (98–128 min) occurred between 2.0 and 12 $R_\odot$. Progressively longer periods manifest at lower heights, that is 0.10–0.15 mHz (111–167 min) at 1.2–2.0 $R_\odot$ and toward 0.08 mHz (208 min) down to 1.1 $R_\odot$. Before interpreting these results, there are many factors to consider. For example, the decrease of the number of OPDs forming below 1.6 $R_\odot$ could lead to less tracks determining longer waiting times between consecutive OPDs and a consequent apparent increase in period. Additionally, at lower heights there are also more entities that might affect our analysis: for example faint tracks resembling IPDs that we did not consider in our analysis plus more complex dynamics related to loop expansion. Nevertheless, we presented one example in which the periodicity of OPDs remained almost unaltered down to $\approx$2.0 $R_\odot$ at about 98–128 min, well within the FOV of COR1.

\section{Discussion} 
\label{sec:discussion}

\begin{figure}[h]
        \centering
         \includegraphics[width=\linewidth]{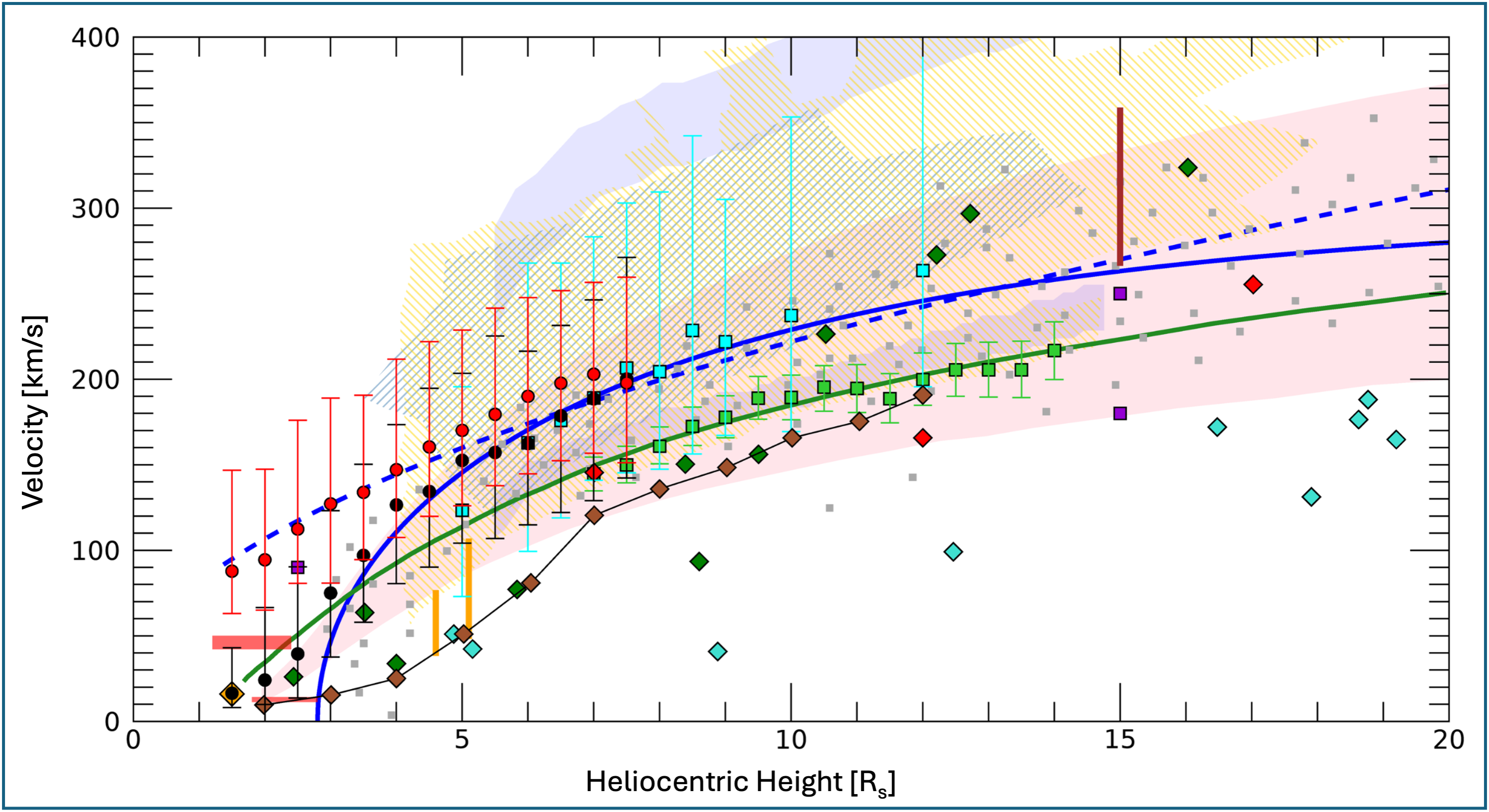}
         \includegraphics[width=\linewidth]{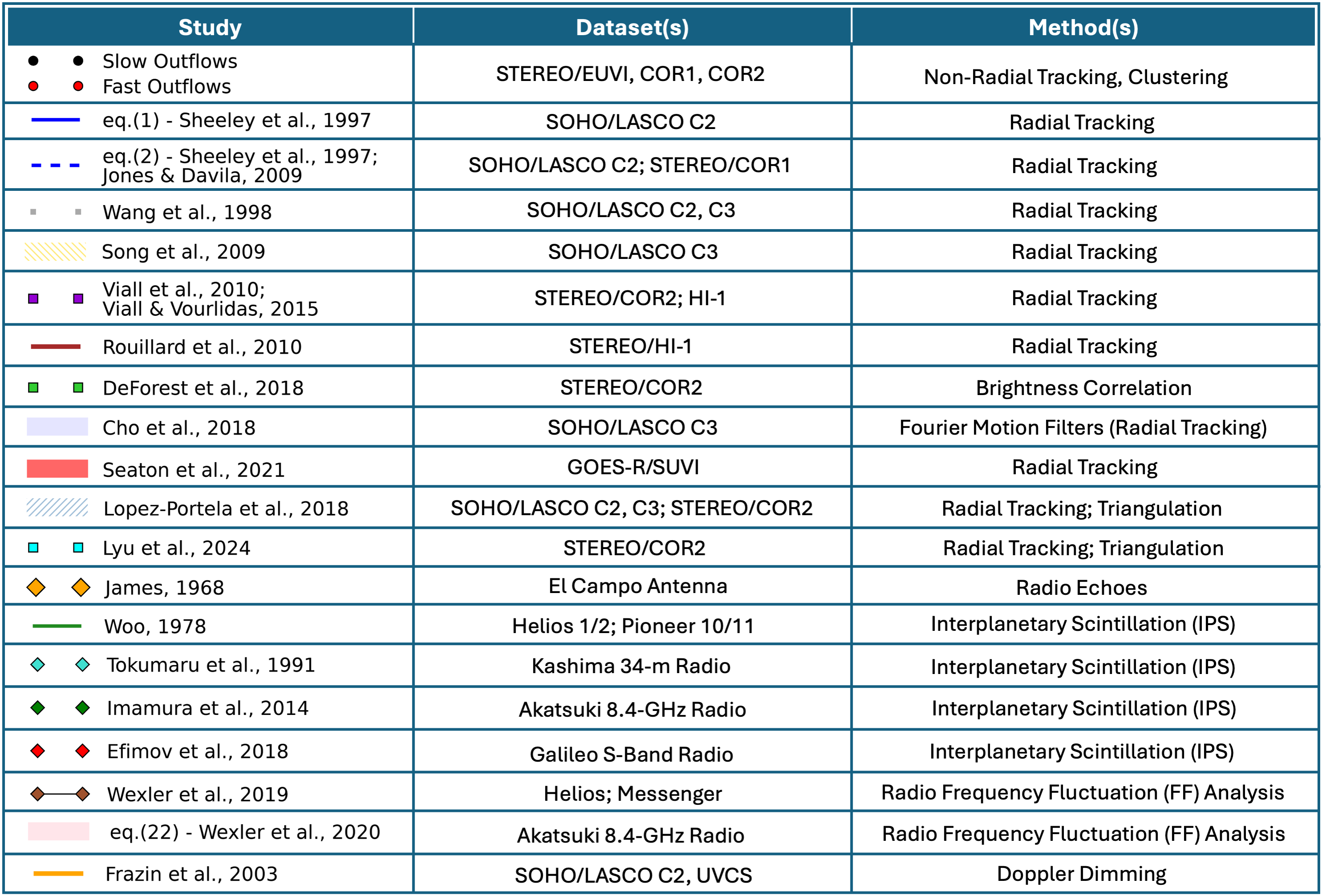}
        \caption{Height-Velocity profiles for slow (black error bars) and fast (red error bars) OPDs compared to estimates and profiles by previous investigations, with the related datasets and methodology summarized in the legend.}
         \label{fig:9_summary_comparison}
    \end{figure}

The analysis of 417 OPDs identified in nonradial Ht-T plots from the STEREO/SECCHI suite shows the occurrence of two main populations that we classified as slow and fast OPDs. To better understand their nature, we collected information from previous works reporting velocity estimates of outflows or bulk coronal plasma at distances from 1.0 to 20 $R_\odot$. The results are summarized in Figure \ref{fig:9_summary_comparison}. A first consideration is that studies that involve the tracking of small-scale outflows show very good agreement above $\approx$5.0 $R_\odot$ with an average speed of $\approx$140 $km/s$ and reaching an average speed of $\approx$300 $km/s$ at 20 $R_\odot$ \citep{Sheeley1997,Wang1998,Jones2009,Viall2010,Viall2015,Rouillard2010,DeForest2018,Cho2018}. Recent investigations using multiple spacecraft observations to triangulate the location of small transients reported radial velocities with a similar height profile but at higher values \citep{Lopez2018,Lyu2024}, remarking how the POS speeds constitute a lower limit for true speeds. Velocity estimates based on Doppler Dimming and radio studies \citep{Tokumaru1991,Frazin2003,Imamura2014,Efimov2018,Wexler2019,Wexler2020} cover a similar range of speeds but extend to lower values: lower than 100 $km/s$ at 5.0 $R_\odot$ and 200 $km/s$ at 20 $R_\odot$. Below 5.0 $R_\odot$, the estimated speed from the tracking of small-scale features shows some discrepancy. \cite{Jones2009} tracked outflows in STEREO/COR1 and reported speeds on the order of 100 $km/s$ down to 1.4 $R_\odot$ and found consistencies with an extrapolation of a velocity profile previously proposed by \citet{Sheeley1997}. On the other hand, recent observations revisiting COR1 observations in the context of nonradial motion of small scale outflows \citep{Alzate2023} and using off-point EUV observations by the SUVI instrument \citep{Seaton2021} have confirmed the occurrence of fast outflows and also revealed the presence of a secondary slower population of outflows. This secondary class of outflows forms as low as 1.2 $R_\odot$ and propagate outward with speeds of tens of $km/s$. In this investigation, the presence of the two OPD populations was suggested by the presence of two main clusters in the velocity profiles shown in Figure \ref{fig:5_velacc_slow_fast}. The same median values and interquartile ranges are reported in Figure \ref{fig:9_summary_comparison} for slow OPDs (black) and fast OPDs (red). Below 3.0 $R_\odot$, we found consistent results between the fast OPDs and previous reports based on white light transient speeds \citep{Jones2009,Viall2015}. For the slow OPDs, the results are consistent with previous solar wind speeds obtained from transients in EUV and WL observations \citep{Alzate2021,Alzate2023,Seaton2021} and deduced from radio studies \citep{Woo1978,Imamura2014,Wexler2019,Wexler2020} down to 1.5 $R_\odot$ \citep{James1968}. Between 3.0 and 4.0 $R_\odot$, there is a general agreement of almost all previous studies in that height range and the velocity profile of the slow OPDs, with the profile of the fast OPDs appearing as an upper limit. Above 4.0 $R_\odot$, both slow and fast OPDs are in good agreement with previously reported speeds from propagating transients \citep{Sheeley1997,Wang1998,Song2009,Viall2010,Rouillard2010,DeForest2018,Lopez2018,Cho2018,Lyu2024} and radio studies \citep{Woo1978,Imamura2014,Efimov2018,Wexler2019,Wexler2020}. Note also that both OPD classes are significantly different from the speed profile of fast wind originating from the sun poles estimated by \cite{Cho2018} that reaches values of $\approx$300 $km/s$ as low as 6.0 $R_\odot$ (light blue band at higher speeds in Figure \ref{fig:9_summary_comparison}).

The observed formation height, speed, and acceleration profiles of the two OPD classes provide some insight into the possible source, release, and acceleration mechanisms of such structures \citep{Viall2020}. Slow OPDs are more often observed near the streamer boundary and originate low in the corona at heights of $\approx$1.6–1.8 $R_\odot$. They slowly move outward and undergo a significant acceleration starting at $\approx$3.0 $R_\odot$. There is no evidence of their formation at this height, suggesting some possible scenarios: (i) blobs occurring in and around the heliospheric current sheet and helmet streamer, closed flux could expand slowly and suddenly be released due to reconnection at $\approx$3.0 $R_\odot$; (ii) plasma is already released at $\approx$1.6–1.8 $R_\odot$, possibly by interchange reconnection, and undergoes a steeper acceleration at $\approx$3.0 $R_\odot$; (iii) for observations on the West limb showing an active region, the speeds below $\approx$3.0 $R_\odot$ could be related to the expansion of active region loops, also showing radial speeds of tens of $km/s$ \citep{Uchida1992,Wang1998,Morgan2013}, which might occur in association with blobs released at the tips of streamers; (iv) the small OPDs population at very low speeds below $\approx$3.0 $R_\odot$, could be related to the streamer motion toward the POS as the sun rotates, which would yield speeds of a few $km/s$, and again higher speeds due to blob release at the tips of streamers; (v) our results are consistent with solar wind speeds deduced from radio studies \citep{Wexler2020} which associate the density variations below and above 2.0 $R_\odot$ to acoustic waves and advected structures, respectively. The fast OPDs on the other hand originate both in the low and middle corona at heights of $\approx$1.6–1.8 $R_\odot$ and $\approx$3.0 $R_\odot$ and both at the streamer boundary and within the streamers. The high speed of the fast OPDs at all heights suggests a likely relation to more impulsive release mechanisms possibly related to interchange reconnection lower in the corona \citep[e. g.,][]{Raouafi2023,Alzate2016} and magnetic reconnection/tearing mode at the tip of streamers \citep{Endeve2004,Endeve2005,Lynch2014,Higginson2018,Reville2020}. After their release, the fast outflow undergoes a small acceleration suggesting a further acceleration mechanism possibly due to wave/turbulence.

The periodic release of OPDs has been associated with mechanisms involving reconnection \citep{Viall2015,Kepko2024}. In this work, we were able to investigate the periodic nature of OPDs down to 1.0 $R_\odot$ and were able to show that periodic brightness variations in the range of 98–128 min can persist down to 2.0 $R_\odot$. Below this height, even though periodicity occurs, its interpretation is hindered by the fact that most of the OPDs form in that region and IPDs might be present as well.

\section{Conclusions} 
\label{sec:conclusions}
The combination of the uninterrupted FOV supplied by the STEREO/SECCHI suite of instruments and the application of advanced image processing techniques has enabled a detailed analysis of OPDs and IPDs during a period of solar minimum. The low solar activity led to a clear identification of streamer profiles from which we were able to define nonradial paths and minimize the effects of nonradial outflow motion and effect due to streamer motion. The properties of the identified OPDs can be summarized as follows:
\begin{itemize}
    \item Based on the velocity profile of 417 unique OPDs we were able to distinguish two classes of OPDs which we refer to as slow and fast OPDs.
    \item Slow OPDs preferentially form at $\approx$1.6 $R_\odot$ closer to the streamer boundary and show speeds of $16.4_{-8.4}^{+26.6}km/s$ at 1.5 $R_\odot$ and accelerate up to $200.1_{-57.9}^{+71.1}km/s$ at 7.5 $R_\odot$.
    \item Fast OPDs preferentially form at $\approx$1.6 $R_\odot$ and at $\approx$3.0 $R_\odot$ both at the streamer boundary and within the streamer. They show speeds of $87.8_{-24.8}^{+59.1}km/s$ at 1.5 $R_\odot$ up to $197.8_{-46.7}^{+61.8}km/s$ at 7.5 $R_\odot$.
    \item IPDs are observed forming at $\approx$1.8 $R_\odot$ with speeds of tens of $km/s$, however, the tracks we were able to identify were mostly concentrated both in location and time in the aftermath of a CME eruption.
    \item The velocity profiles of slow OPDs for heliocentric height below 3.0 $R_\odot$ show good agreement with speeds more closely related to the bulk solar wind obtained via Doppler dimming and interplanetary scintillation.
    \item We presented one example in which we were able to show that periodic brightness variations related to OPDs remained in the range of 98–128 min down to $\approx$2.0 $R_\odot$, well within the FOV of COR1.
\end{itemize}

The OPDs presented in this work occurred during a period of solar minimum when the magnetic topology was simple and dipolar with little magnetic complexity in the global magnetic field.  A current effort is underway where we are focused on magnetically complex time periods throughout Solar Cycle 24. Analyzing several time periods during an entire solar cycle is a more general and stringent test on solar wind theories and will address questions regarding the ambient structures in the solar wind in terms of occurrence location, occurrence rate, formation height, and formation mechanism.

The results of this work have also highlighted that without the separation of OPDs in the slow and fast classes there is inconsistency between the speeds obtained by tracking small scale OPDs and the one more closely related to the bulk solar wind (e. g., Doppler dimming and IPS) below 3 $R_\odot$. Future observations from the COronal Diagnostic EXperiment \citep[CODEX;][]{Cho2017,Cho2020} could shed new light on this duality comparing the speed of small OPDs with the global velocity maps that this new mission will provide.

In our current investigation we limited our analysis to heights below 8.0 $R_\odot$. A more extensive analysis involving observations at larger heights would be very valuable in understating the evolution of slow and fast OPDs beyond 8.0 $R_\odot$. In this regard, the upcoming Polarimeter to UNify the Corona and Heliosphere \citep[PUNCH;][]{DeForest2022} mission will provide the opportunity to trace these OPDs close to 1 AU.

Finally, the clear distinction between slow and fast OPDs was enabled by the observations below 3.0 $R_\odot$, remarking the fundamental importance of this region. Routine observations with higher resolution and multi-point extended FOV involving widely overlapping EUV and white light observations in this region would really benefit the study of the connection among the low, middle, and high corona \citep{West2023}.\\



N.A. acknowledges support from NASA ROSES through HGI grant No. 80NSSC20K1070 and PSP-GI grant No. 80NSSC21K1945. The work of S.D.M. was supported under the ECIP grant No. 80NSSC21K0459 and PSP-GI grant No. 80NSSC21K1945.  H.M. acknowledges PSP-GI grant No. 80NSSC21K1945.  N.V. is supported by the Goddard Space Flight Center Heliophysics Internal Scientist Funding Model (ISFM; competitive work package). A.V. is supported by  NASA grant 80NSSC22K0970.  The tomography maps are available at \url{https://solarphysics.aber.ac.uk/Archives/tomography/}. The authors also thank the anonymous reviewer for the useful comments that helped improve the quality of this paper.


\vspace{5mm}
\facilities{STEREO (EUVI, COR1 and COR2). The tomography maps were produced using SuperComputing Wales.}

\software{The spectral analysis code used in this work is freely available on the Zenodo platform \citep{DiMatteo2020}}




\appendix

\section{Revisited kinematics of slow OPDs}
\begin{figure}[h!]
        \centering
         \includegraphics[width=0.8\linewidth]{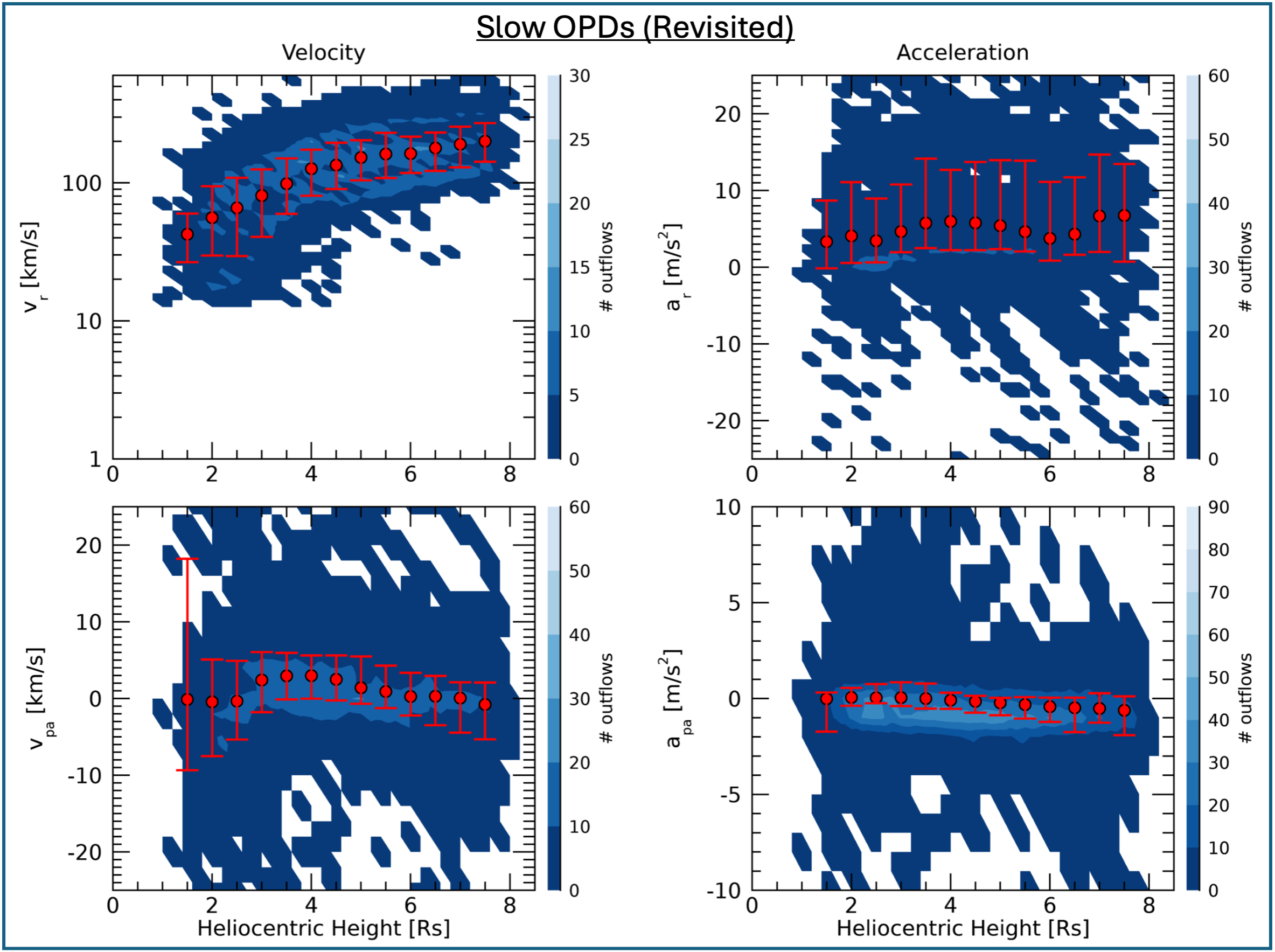}
        \caption{Revisited velocity and acceleration profiles for slow OPDs. Top row shows the profile of $v_r$ and $a_r$, while the bottom row shows the $v_{pa}$ and $a_{pa}$ profile. The color scale indicates the number of OPDs while the errorbars indicate the median value (red dot) and interquartile range (red bars) at height bins of 0.5 $R_\odot$.}
         \label{fig:10_revisited_slow}
    \end{figure}

We take into account the possible presence of artifacts due to solar rotation in the radial velocity profile of slow OPDs. We remove from our analysis the portion of the unique tracks with speeds lower than $\approx$15 $km/s$. Figure \ref{fig:10_revisited_slow} shows the revisited kinematic analysis: the slow OPDs have a lower initial radial speed of $42.4_{-15.8}^{+17.5}km/s$ at 1.5 $R_\odot$ and accelerate up to $200.1_{-57.9}^{+71.1}km/s$ at 7.5 $R_\odot$. The radial acceleration profile shows a gradual increase, with some fluctuations, from $3.3_{-3.3}^{+5.4}m/s^2$ at 1.5 $R_\odot$ to 
$6.7_{-6.0}^{+6.7}m/s^2$ at 7.5 $R_\odot$. The median $v_{pa}$ shows slight negative values below 2.5 $R_\odot$ after which it assumes positive values (motion toward the center of the streamer) peaking with a speed of $3.0_{-3.0}^{+2.6}km/s$ at 4.0 $R_\odot$ and decreasing to almost null values at 7.5 $R_\odot$. The median $a_{pa}$ values showed accordingly slight positive and negative values respectively below and above 4.0 $R_\odot$. Note also that the formation location results remain largely unchanged from the ones reported in Figure \ref{fig:6_formation_height}a.

\bibliography{Alzate_et-al}{}

\begin{thebibliography}{}
\expandafter\ifx\csname natexlab\endcsname\relax\def\natexlab#1{#1}\fi
\providecommand{\url}[1]{\href{#1}{#1}}
\providecommand{\dodoi}[1]{doi:~\href{http://doi.org/#1}{\nolinkurl{#1}}}
\providecommand{\doeprint}[1]{\href{http://ascl.net/#1}{\nolinkurl{http://ascl.net/#1}}}
\providecommand{\doarXiv}[1]{\href{https://arxiv.org/abs/#1}{\nolinkurl{https://arxiv.org/abs/#1}}}

\bibitem[{Alzate {et~al.}(2024)Alzate, Di~Matteo, Morgan, Viall, \&
  Vourlidas}]{Alzate2024}
Alzate, N., Di~Matteo, S., Morgan, H., Viall, N.~M., \& Vourlidas, A. 2024,
  Supporting Movie for ``Connecting the Low to High Corona: Propagating
  Disturbances as Tracers of the Near-Sun Solar Wind'' by Alzate et al. (2024),
  1.0, \dodoi{10.5281/zenodo.11211569}

\bibitem[{{Alzate} \& {Morgan}(2016)}]{Alzate2016}
{Alzate}, N., \& {Morgan}, H. 2016, \apj, 823, 129,
  \dodoi{10.3847/0004-637X/823/2/129}

\bibitem[{{Alzate} {et~al.}(2023){Alzate}, {Morgan}, \& {Di
  Matteo}}]{Alzate2023}
{Alzate}, N., {Morgan}, H., \& {Di Matteo}, S. 2023, \apj, 945, 116,
  \dodoi{10.3847/1538-4357/acba08}

\bibitem[{{Alzate} {et~al.}(2021){Alzate}, {Morgan}, {Viall}, \&
  {Vourlidas}}]{Alzate2021}
{Alzate}, N., {Morgan}, H., {Viall}, N., \& {Vourlidas}, A. 2021, \apj, 919,
  98, \dodoi{10.3847/1538-4357/ac10ca}

\bibitem[{{Antiochos} {et~al.}(2011){Antiochos}, {Miki{\'c}}, {Titov},
  {Lionello}, \& {Linker}}]{Antiochos2011}
{Antiochos}, S.~K., {Miki{\'c}}, Z., {Titov}, V.~S., {Lionello}, R., \&
  {Linker}, J.~A. 2011, \apj, 731, 112, \dodoi{10.1088/0004-637X/731/2/112}

\bibitem[{{Antonucci} {et~al.}(2020){Antonucci}, {Romoli}, {Andretta},
  {Fineschi}, {Heinzel}, {Moses}, {Naletto}, {Nicolini}, {Spadaro}, {Teriaca},
  {Berlicki}, {Capobianco}, {Crescenzio}, {Da Deppo}, {Focardi}, {Frassetto},
  {Heerlein}, {Landini}, {Magli}, {Marco Malvezzi}, {Massone}, {Melich},
  {Nicolosi}, {Noci}, {Pancrazzi}, {Pelizzo}, {Poletto}, {Sasso},
  {Sch{\"u}hle}, {Solanki}, {Strachan}, {Susino}, {Tondello}, {Uslenghi},
  {Woch}, {Abbo}, {Bemporad}, {Casti}, {Dolei}, {Grimani}, {Messerotti},
  {Ricci}, {Straus}, {Telloni}, {Zuppella}, {Auch{\`e}re}, {Bruno},
  {Ciaravella}, {Corso}, {Alvarez Copano}, {Aznar Cuadrado}, {D'Amicis},
  {Enge}, {Gravina}, {Jej{\v{c}}i{\v{c}}}, {Lamy}, {Lanzafame}, {Meierdierks},
  {Papagiannaki}, {Peter}, {Fernandez Rico}, {Giday Sertsu}, {Staub},
  {Tsinganos}, {Velli}, {Ventura}, {Verroi}, {Vial}, {Vives}, {Volpicelli},
  {Werner}, {Zerr}, {Negri}, {Castronuovo}, {Gabrielli}, {Bertacin},
  {Carpentiero}, {Natalucci}, {Marliani}, {Cesa}, {Laget}, {Morea},
  {Pieraccini}, {Radaelli}, {Sandri}, {Sarra}, {Cesare}, {Del Forno}, {Massa},
  {Montabone}, {Mottini}, {Quattropani}, {Schillaci}, {Boccardo}, {Brando},
  {Pandi}, {Baietto}, {Bertone}, {Alvarez-Herrero}, {Garc{\'\i}a Parejo},
  {Cebollero}, {Amoruso}, \& {Centonze}}]{Antonucci2020}
{Antonucci}, E., {Romoli}, M., {Andretta}, V., {et~al.} 2020, \aap, 642, A10,
  \dodoi{10.1051/0004-6361/201935338}

\bibitem[{{Boe} {et~al.}(2020){Boe}, {Habbal}, \& {Druckm{\"u}ller}}]{Boe2020}
{Boe}, B., {Habbal}, S., \& {Druckm{\"u}ller}, M. 2020, \apj, 895, 123,
  \dodoi{10.3847/1538-4357/ab8ae6}

\bibitem[{{Byrne} {et~al.}(2013){Byrne}, {Long}, {Gallagher}, {Bloomfield},
  {Maloney}, {McAteer}, {Morgan}, \& {Habbal}}]{Byrne2013}
{Byrne}, J.~P., {Long}, D.~M., {Gallagher}, P.~T., {et~al.} 2013, \aap, 557,
  A96, \dodoi{10.1051/0004-6361/201321223}

\bibitem[{{Cho} {et~al.}(2018){Cho}, {Moon}, {Nakariakov}, {Bong}, {Lee},
  {Song}, {Lee}, \& {Cho}}]{Cho2018}
{Cho}, I.-H., {Moon}, Y.-J., {Nakariakov}, V.~M., {et~al.} 2018, \prl, 121,
  075101, \dodoi{10.1103/PhysRevLett.121.075101}

\bibitem[{{Cho} {et~al.}(2017){Cho}, {Bong}, {Choi}, {Yang}, {Kim}, {Baek},
  {Park}, {Lim}, {Kim}, {Kim}, {Kim}, {Park}, {Clarke}, {Davila}, {Gopalswamy},
  {Nakariakov}, {Li}, \& {Pinto}}]{Cho2017}
{Cho}, K.~S., {Bong}, S.~C., {Choi}, S., {et~al.} 2017, Journal of Korean
  Astronomical Society, 50, 139, \dodoi{10.5303/JKAS.2017.50.5.139}

\bibitem[{{Cho} {et~al.}(2020){Cho}, {Yang}, {Lee}, {Bong}, {Kim}, {Choi},
  {Park}, {Cho}, {Baek}, {Kim}, \& {Park}}]{Cho2020}
{Cho}, K.~S., {Yang}, H., {Lee}, J.~O., {et~al.} 2020, Journal of Korean
  Astronomical Society, 53, 87, \dodoi{10.5303/JKAS.2020.53.4.87}

\bibitem[{{Claudepierre} {et~al.}(2009){Claudepierre}, {Wiltberger},
  {Elkington}, {Lotko}, \& {Hudson}}]{Claudepierre2009}
{Claudepierre}, S.~G., {Wiltberger}, M., {Elkington}, S.~R., {Lotko}, W., \&
  {Hudson}, M.~K. 2009, \grl, 36, L13101, \dodoi{10.1029/2009GL039045}

\bibitem[{{Crooker} {et~al.}(2004){Crooker}, {Huang}, {Lamassa}, {Larson},
  {Kahler}, \& {Spence}}]{Crooker2004}
{Crooker}, N.~U., {Huang}, C.~L., {Lamassa}, S.~M., {et~al.} 2004, Journal of
  Geophysical Research (Space Physics), 109, A03107,
  \dodoi{10.1029/2003JA010170}

\bibitem[{{DeForest} {et~al.}(2022){DeForest}, {Killough}, {Gibson}, {Henry},
  {Case}, {Beasley}, {Laurent}, {Colaninno}, {Waltham}, \& {Punch Science
  Team}}]{DeForest2022}
{DeForest}, C., {Killough}, R., {Gibson}, S., {et~al.} 2022, in 2022 IEEE
  Aerospace Conference, 1--11, \dodoi{10.1109/AERO53065.2022.9843340}

\bibitem[{{DeForest} {et~al.}(2018){DeForest}, {Howard}, {Velli}, {Viall}, \&
  {Vourlidas}}]{DeForest2018}
{DeForest}, C.~E., {Howard}, R.~A., {Velli}, M., {Viall}, N., \& {Vourlidas},
  A. 2018, \apj, 862, 18, \dodoi{10.3847/1538-4357/aac8e3}

\bibitem[{{DeForest} {et~al.}(2016){DeForest}, {Matthaeus}, {Viall}, \&
  {Cranmer}}]{DeForest2016}
{DeForest}, C.~E., {Matthaeus}, W.~H., {Viall}, N.~M., \& {Cranmer}, S.~R.
  2016, \apj, 828, 66, \dodoi{10.3847/0004-637X/828/2/66}

\bibitem[{Di~Matteo {et~al.}(2020)Di~Matteo, Viall, \& Kepko}]{DiMatteo2020}
Di~Matteo, S., Viall, N.~M., \& Kepko, L. 2020, {SPD}\_{MTM}: a spectral
  analysis tool for the {SPEDAS} framework, v1.0,
  \dodoi{10.5281/zenodo.3703168}

\bibitem[{{Di Matteo} {et~al.}(2021){Di Matteo}, {Viall}, \&
  {Kepko}}]{DiMatteo2021}
{Di Matteo}, S., {Viall}, N.~M., \& {Kepko}, L. 2021, Journal of Geophysical
  Research (Space Physics), 126, e28748, \dodoi{10.1029/2020JA028748}

\bibitem[{{Di Matteo} {et~al.}(2019){Di Matteo}, {Viall}, {Kepko}, {Wallace},
  {Arge}, \& {MacNeice}}]{DiMatteo2019}
{Di Matteo}, S., {Viall}, N.~M., {Kepko}, L., {et~al.} 2019, Journal of
  Geophysical Research (Space Physics), 124, 837, \dodoi{10.1029/2018JA026182}

\bibitem[{{Di Matteo} \& {Villante}(2017)}]{DiMatteo2017}
{Di Matteo}, S., \& {Villante}, U. 2017, Journal of Geophysical Research (Space
  Physics), 122, 4905, \dodoi{10.1002/2017JA023936}

\bibitem[{{Di Matteo} {et~al.}(2022){Di Matteo}, {Villante}, {Viall}, {Kepko},
  \& {Wallace}}]{DiMatteo2022}
{Di Matteo}, S., {Villante}, U., {Viall}, N., {Kepko}, L., \& {Wallace}, S.
  2022, Journal of Geophysical Research (Space Physics), 127, e30144,
  \dodoi{10.1029/2021JA03014410.1002/essoar.10508935.1}

\bibitem[{{Edmondson}(2012)}]{Edmondson2012}
{Edmondson}, J.~K. 2012, \ssr, 172, 209, \dodoi{10.1007/s11214-011-9767-y}

\bibitem[{{Efimov} {et~al.}(2018){Efimov}, {Lukanina}, {Chashei}, {Bird},
  {P{\"a}tzold}, \& {Wexler}}]{Efimov2018}
{Efimov}, A.~I., {Lukanina}, L.~A., {Chashei}, I.~V., {et~al.} 2018, Cosmic
  Research, 56, 405, \dodoi{10.1134/S0010952518060023}

\bibitem[{{Einaudi} {et~al.}(1999){Einaudi}, {Boncinelli}, {Dahlburg}, \&
  {Karpen}}]{Einaudi1999}
{Einaudi}, G., {Boncinelli}, P., {Dahlburg}, R.~B., \& {Karpen}, J.~T. 1999,
  \jgr, 104, 521, \dodoi{10.1029/98JA02394}

\bibitem[{{Einaudi} {et~al.}(2001){Einaudi}, {Chibbaro}, {Dahlburg}, \&
  {Velli}}]{Einaudi2001}
{Einaudi}, G., {Chibbaro}, S., {Dahlburg}, R.~B., \& {Velli}, M. 2001, \apj,
  547, 1167, \dodoi{10.1086/318400}

\bibitem[{Endeve {et~al.}(2004)Endeve, Holzer, \& Leer}]{Endeve2004}
Endeve, E., Holzer, T.~E., \& Leer, E. 2004, The Astrophysical Journal, 603,
  307–321, \dodoi{10.1086/381239}

\bibitem[{Endeve {et~al.}(2005)Endeve, Lie‐Svendsen, Hansteen, \&
  Leer}]{Endeve2005}
Endeve, E., Lie‐Svendsen, O., Hansteen, V.~H., \& Leer, E. 2005, The
  Astrophysical Journal, 624, 402–413, \dodoi{10.1086/428938}

\bibitem[{{Everitt}(1993)}]{Everitt1993}
{Everitt}, B.~S. 1993, {Cluster Analysis} (New York: Halsted Press)

\bibitem[{{Fisk} \& {Schwadron}(2001)}]{Fisk2001}
{Fisk}, L.~A., \& {Schwadron}, N.~A. 2001, \apj, 560, 425,
  \dodoi{10.1086/322503}

\bibitem[{{Fox} {et~al.}(2016){Fox}, {Velli}, {Bale}, {Decker}, {Driesman},
  {Howard}, {Kasper}, {Kinnison}, {Kusterer}, {Lario}, {Lockwood}, {McComas},
  {Raouafi}, \& {Szabo}}]{Fox2016}
{Fox}, N.~J., {Velli}, M.~C., {Bale}, S.~D., {et~al.} 2016, \ssr, 204, 7,
  \dodoi{10.1007/s11214-015-0211-6}

\bibitem[{{Frazin} {et~al.}(2003){Frazin}, {Cranmer}, \& {Kohl}}]{Frazin2003}
{Frazin}, R.~A., {Cranmer}, S.~R., \& {Kohl}, J.~L. 2003, \apj, 597, 1145,
  \dodoi{10.1086/378558}

\bibitem[{{Harrison} {et~al.}(2009){Harrison}, {Davis}, \&
  {Davies}}]{Harrison2009}
{Harrison}, R.~A., {Davis}, C.~J., \& {Davies}, J.~A. 2009, \solphys, 259, 277,
  \dodoi{10.1007/s11207-009-9417-7}

\bibitem[{{Hartinger} {et~al.}(2014){Hartinger}, {Welling}, {Viall}, {Moldwin},
  \& {Ridley}}]{Hartinger2014}
{Hartinger}, M.~D., {Welling}, D., {Viall}, N.~M., {Moldwin}, M.~B., \&
  {Ridley}, A. 2014, Journal of Geophysical Research (Space Physics), 119,
  8212, \dodoi{10.1002/2014JA020401}

\bibitem[{{Hess} \& {Wang}(2017)}]{Hess2017}
{Hess}, P., \& {Wang}, Y.~M. 2017, \apj, 850, 6,
  \dodoi{10.3847/1538-4357/aa921d}

\bibitem[{{Higginson} {et~al.}(2017){Higginson}, {Antiochos}, {DeVore},
  {Wyper}, \& {Zurbuchen}}]{Higginson2017}
{Higginson}, A.~K., {Antiochos}, S.~K., {DeVore}, C.~R., {Wyper}, P.~F., \&
  {Zurbuchen}, T.~H. 2017, \apj, 837, 113, \dodoi{10.3847/1538-4357/837/2/113}

\bibitem[{{Higginson} \& {Lynch}(2018)}]{Higginson2018}
{Higginson}, A.~K., \& {Lynch}, B.~J. 2018, \apj, 859, 6,
  \dodoi{10.3847/1538-4357/aabc08}

\bibitem[{{Howard} {et~al.}(2008){Howard}, {Moses}, {Vourlidas}, {Newmark},
  {Socker}, {Plunkett}, {Korendyke}, {Cook}, {Hurley}, {Davila}, {Thompson},
  {St Cyr}, {Mentzell}, {Mehalick}, {Lemen}, {Wuelser}, {Duncan}, {Tarbell},
  {Wolfson}, {Moore}, {Harrison}, {Waltham}, {Lang}, {Davis}, {Eyles},
  {Mapson-Menard}, {Simnett}, {Halain}, {Defise}, {Mazy}, {Rochus}, {Mercier},
  {Ravet}, {Delmotte}, {Auchere}, {Delaboudiniere}, {Bothmer}, {Deutsch},
  {Wang}, {Rich}, {Cooper}, {Stephens}, {Maahs}, {Baugh}, {McMullin}, \&
  {Carter}}]{Howard2008}
{Howard}, R.~A., {Moses}, J.~D., {Vourlidas}, A., {et~al.} 2008, \ssr, 136, 67,
  \dodoi{10.1007/s11214-008-9341-4}

\bibitem[{{Howard} {et~al.}(2019){Howard}, {Vourlidas}, {Bothmer}, {Colaninno},
  {DeForest}, {Gallagher}, {Hall}, {Hess}, {Higginson}, {Korendyke},
  {Kouloumvakos}, {Lamy}, {Liewer}, {Linker}, {Linton}, {Penteado}, {Plunkett},
  {Poirier}, {Raouafi}, {Rich}, {Rochus}, {Rouillard}, {Socker}, {Stenborg},
  {Thernisien}, \& {Viall}}]{Howard2019}
{Howard}, R.~A., {Vourlidas}, A., {Bothmer}, V., {et~al.} 2019, \nat, 576, 232,
  \dodoi{10.1038/s41586-019-1807-x}

\bibitem[{{Imamura} {et~al.}(2014){Imamura}, {Tokumaru}, {Isobe}, {Shiota},
  {Ando}, {Miyamoto}, {Toda}, {H{\"a}usler}, {P{\"a}tzold}, {Nabatov}, {Asai},
  {Yaji}, {Yamada}, \& {Nakamura}}]{Imamura2014}
{Imamura}, T., {Tokumaru}, M., {Isobe}, H., {et~al.} 2014, \apj, 788, 117,
  \dodoi{10.1088/0004-637X/788/2/117}

\bibitem[{{James}(1968)}]{James1968}
{James}, J.~C. 1968, {Radar astronomy}, ed. J.~V. {Evans} \& T.~{Hagfors}

\bibitem[{{Jones} \& {Davila}(2009)}]{Jones2009}
{Jones}, S.~I., \& {Davila}, J.~M. 2009, \apj, 701, 1906,
  \dodoi{10.1088/0004-637X/701/2/1906}

\bibitem[{{Kaiser} {et~al.}(2008){Kaiser}, {Kucera}, {Davila}, {St. Cyr},
  {Guhathakurta}, \& {Christian}}]{Kaiser2008}
{Kaiser}, M.~L., {Kucera}, T.~A., {Davila}, J.~M., {et~al.} 2008, \ssr, 136, 5,
  \dodoi{10.1007/s11214-007-9277-0}

\bibitem[{{Kepko} \& {Spence}(2003)}]{Kepko2003}
{Kepko}, L., \& {Spence}, H.~E. 2003, Journal of Geophysical Research (Space
  Physics), 108, 1257, \dodoi{10.1029/2002JA009676}

\bibitem[{{Kepko} {et~al.}(2002){Kepko}, {Spence}, \& {Singer}}]{Kepko2002}
{Kepko}, L., {Spence}, H.~E., \& {Singer}, H.~J. 2002, \grl, 29, 1197,
  \dodoi{10.1029/2001GL014405}

\bibitem[{{Kepko} \& {Viall}(2019)}]{Kepko2019}
{Kepko}, L., \& {Viall}, N.~M. 2019, Journal of Geophysical Research (Space
  Physics), 124, 7722, \dodoi{10.1029/2019JA026962}

\bibitem[{{Kepko} {et~al.}(2016){Kepko}, {Viall}, {Antiochos}, {Lepri},
  {Kasper}, \& {Weberg}}]{Kepko2016}
{Kepko}, L., {Viall}, N.~M., {Antiochos}, S.~K., {et~al.} 2016, \grl, 43, 4089,
  \dodoi{10.1002/2016GL068607}

\bibitem[{{Kepko} {et~al.}(2024){Kepko}, {Viall}, \& {Di Matteo}}]{Kepko2024}
{Kepko}, L., {Viall}, N.~M., \& {Di Matteo}, S. 2024, Journal of Geophysical
  Research (Space Physics), 129, e2023JA031403, \dodoi{10.1029/2023JA031403}

\bibitem[{{Kumar} {et~al.}(2023){Kumar}, {Karpen}, {Antiochos}, {DeVore},
  {Wyper}, \& {Cho}}]{Kumar2023}
{Kumar}, P., {Karpen}, J.~T., {Antiochos}, S.~K., {et~al.} 2023, \apj, 943,
  156, \dodoi{10.3847/1538-4357/acaea4}

\bibitem[{{Kumar} {et~al.}(2022){Kumar}, {Karpen}, {Uritsky}, {Deforest},
  {Raouafi}, \& {Richard DeVore}}]{Kumar2022}
{Kumar}, P., {Karpen}, J.~T., {Uritsky}, V.~M., {et~al.} 2022, \apj, 933, 21,
  \dodoi{10.3847/1538-4357/ac6c24}

\bibitem[{{Lapenta} \& {Knoll}(2005)}]{Lapenta2005}
{Lapenta}, G., \& {Knoll}, D.~A. 2005, \apj, 624, 1049, \dodoi{10.1086/429262}

\bibitem[{{Liu} {et~al.}(2020){Liu}, {Wang}, {Wimmer-Schweingruber}, {Krucker},
  \& {Mason}}]{Liu2020}
{Liu}, Z., {Wang}, L., {Wimmer-Schweingruber}, R.~F., {Krucker}, S., \&
  {Mason}, G.~M. 2020, Journal of Geophysical Research (Space Physics), 125,
  e28702, \dodoi{10.1029/2020JA028702}

\bibitem[{{L{\'o}pez-Portela} {et~al.}(2018){L{\'o}pez-Portela}, {Panasenco},
  {Blanco-Cano}, \& {Stenborg}}]{Lopez2018}
{L{\'o}pez-Portela}, C., {Panasenco}, O., {Blanco-Cano}, X., \& {Stenborg}, G.
  2018, \solphys, 293, 99, \dodoi{10.1007/s11207-018-1315-4}

\bibitem[{Lynch {et~al.}(2014)Lynch, Edmondson, \& Li}]{Lynch2014}
Lynch, B.~J., Edmondson, J.~K., \& Li, Y. 2014, Solar Physics, 289,
  3043–3058, \dodoi{10.1007/s11207-014-0506-x}

\bibitem[{{Lyu} {et~al.}(2024){Lyu}, {Wang}, {Li}, {Zhang}, \& {Liu}}]{Lyu2024}
{Lyu}, S., {Wang}, Y., {Li}, X., {Zhang}, Q., \& {Liu}, J. 2024, \apj, 962,
  170, \dodoi{10.3847/1538-4357/ad1dd5}

\bibitem[{{Mason} {et~al.}(2019){Mason}, {Antiochos}, \& {Viall}}]{Mason2019}
{Mason}, E.~I., {Antiochos}, S.~K., \& {Viall}, N.~M. 2019, \apjl, 874, L33,
  \dodoi{10.3847/2041-8213/ab0c5d}

\bibitem[{{Morgan} {et~al.}(2006){Morgan}, {Habbal}, \& {Woo}}]{Morgan2006}
{Morgan}, H., {Habbal}, S.~R., \& {Woo}, R. 2006, \solphys, 236, 263,
  \dodoi{10.1007/s11207-006-0113-6}

\bibitem[{{Morgan} {et~al.}(2013){Morgan}, {Jeska}, \& {Leonard}}]{Morgan2013}
{Morgan}, H., {Jeska}, L., \& {Leonard}, D. 2013, \apjs, 206, 19,
  \dodoi{10.1088/0067-0049/206/2/19}

\bibitem[{{Poirier} {et~al.}(2023){Poirier}, {R{\'e}ville}, {Rouillard},
  {Kouloumvakos}, \& {Valette}}]{Poirier2023}
{Poirier}, N., {R{\'e}ville}, V., {Rouillard}, A.~P., {Kouloumvakos}, A., \&
  {Valette}, E. 2023, \aap, 677, A108, \dodoi{10.1051/0004-6361/202347146}

\bibitem[{{Raouafi} \& {Stenborg}(2014)}]{Raouafi2014}
{Raouafi}, N.~E., \& {Stenborg}, G. 2014, \apj, 787, 118,
  \dodoi{10.1088/0004-637X/787/2/118}

\bibitem[{{Raouafi} {et~al.}(2016){Raouafi}, {Patsourakos}, {Pariat}, {Young},
  {Sterling}, {Savcheva}, {Shimojo}, {Moreno-Insertis}, {DeVore}, {Archontis},
  {T{\"o}r{\"o}k}, {Mason}, {Curdt}, {Meyer}, {Dalmasse}, \&
  {Matsui}}]{Raouafi2016}
{Raouafi}, N.~E., {Patsourakos}, S., {Pariat}, E., {et~al.} 2016, \ssr, 201, 1,
  \dodoi{10.1007/s11214-016-0260-5}

\bibitem[{{Raouafi} {et~al.}(2023){Raouafi}, {Stenborg}, {Seaton}, {Wang},
  {Wang}, {DeForest}, {Bale}, {Drake}, {Uritsky}, {Karpen}, {DeVore},
  {Sterling}, {Horbury}, {Harra}, {Bourouaine}, {Kasper}, {Kumar}, {Phan}, \&
  {Velli}}]{Raouafi2023}
{Raouafi}, N.~E., {Stenborg}, G., {Seaton}, D.~B., {et~al.} 2023, \apj, 945,
  28, \dodoi{10.3847/1538-4357/acaf6c}

\bibitem[{{R{\'e}ville} {et~al.}(2020){R{\'e}ville}, {Velli}, {Rouillard},
  {Lavraud}, {Tenerani}, {Shi}, \& {Strugarek}}]{Reville2020}
{R{\'e}ville}, V., {Velli}, M., {Rouillard}, A.~P., {et~al.} 2020, \apjl, 895,
  L20, \dodoi{10.3847/2041-8213/ab911d}

\bibitem[{{Rouillard} {et~al.}(2009){Rouillard}, {Savani}, {Davies}, {Lavraud},
  {Forsyth}, {Morley}, {Opitz}, {Sheeley}, {Burlaga}, {Sauvaud}, {Simunac},
  {Luhmann}, {Galvin}, {Crothers}, {Davis}, {Harrison}, {Lockwood}, {Eyles},
  {Bewsher}, \& {Brown}}]{Rouillard2009}
{Rouillard}, A.~P., {Savani}, N.~P., {Davies}, J.~A., {et~al.} 2009, \solphys,
  256, 307, \dodoi{10.1007/s11207-009-9329-6}

\bibitem[{{Rouillard} {et~al.}(2010){Rouillard}, {Davies}, {Lavraud},
  {Forsyth}, {Savani}, {Bewsher}, {Brown}, {Sheeley}, {Davis}, {Harrison},
  {Howard}, {Vourlidas}, {Lockwood}, {Crothers}, \& {Eyles}}]{Rouillard2010}
{Rouillard}, A.~P., {Davies}, J.~A., {Lavraud}, B., {et~al.} 2010, Journal of
  Geophysical Research (Space Physics), 115, A04103,
  \dodoi{10.1029/2009JA014471}

\bibitem[{{Rouillard} {et~al.}(2011){Rouillard}, {Sheeley}, {Cooper}, {Davies},
  {Lavraud}, {Kilpua}, {Skoug}, {Steinberg}, {Szabo}, {Opitz}, \&
  {Sauvaud}}]{Rouillard2011}
{Rouillard}, A.~P., {Sheeley}, N.~R., J., {Cooper}, T.~J., {et~al.} 2011, \apj,
  734, 7, \dodoi{10.1088/0004-637X/734/1/7}

\bibitem[{{Seaton} {et~al.}(2021){Seaton}, {Hughes}, {Tadikonda}, {Caspi},
  {DeForest}, {Krimchansky}, {Hurlburt}, {Seguin}, \& {Slater}}]{Seaton2021}
{Seaton}, D.~B., {Hughes}, J.~M., {Tadikonda}, S.~K., {et~al.} 2021, Nature
  Astronomy, 5, 1029, \dodoi{10.1038/s41550-021-01427-8}

\bibitem[{{Sheeley} {et~al.}(2001){Sheeley}, {Knudson}, \&
  {Wang}}]{Sheeley2001}
{Sheeley}, N.~R., J., {Knudson}, T.~N., \& {Wang}, Y.~M. 2001, \apjl, 546,
  L131, \dodoi{10.1086/318873}

\bibitem[{{Sheeley} {et~al.}(2009){Sheeley}, {Lee}, {Casto}, {Wang}, \&
  {Rich}}]{Sheeley2009}
{Sheeley}, N.~R., J., {Lee}, D.~D.~H., {Casto}, K.~P., {Wang}, Y.~M., \&
  {Rich}, N.~B. 2009, \apj, 694, 1471, \dodoi{10.1088/0004-637X/694/2/1471}

\bibitem[{{Sheeley} \& {Rouillard}(2010)}]{Sheeley2010}
{Sheeley}, N.~R., J., \& {Rouillard}, A.~P. 2010, \apj, 715, 300,
  \dodoi{10.1088/0004-637X/715/1/300}

\bibitem[{{Sheeley} \& {Wang}(2002)}]{Sheeley2002}
{Sheeley}, N.~R., J., \& {Wang}, Y.~M. 2002, \apj, 579, 874,
  \dodoi{10.1086/342923}

\bibitem[{{Sheeley} \& {Wang}(2007)}]{Sheeley2007}
---. 2007, \apj, 655, 1142, \dodoi{10.1086/510323}

\bibitem[{{Sheeley} \& {Wang}(2014)}]{Sheeley2014}
---. 2014, \apj, 797, 10, \dodoi{10.1088/0004-637X/797/1/10}

\bibitem[{{Sheeley} {et~al.}(1997){Sheeley}, {Wang}, {Hawley}, {Brueckner},
  {Dere}, {Howard}, {Koomen}, {Korendyke}, {Michels}, {Paswaters}, {Socker},
  {St. Cyr}, {Wang}, {Lamy}, {Llebaria}, {Schwenn}, {Simnett}, {Plunkett}, \&
  {Biesecker}}]{Sheeley1997}
{Sheeley}, N.~R., {Wang}, Y.~M., {Hawley}, S.~H., {et~al.} 1997, \apj, 484,
  472, \dodoi{10.1086/304338}

\bibitem[{{Song} {et~al.}(2009){Song}, {Chen}, {Liu}, {Feng}, \&
  {Xia}}]{Song2009}
{Song}, H.~Q., {Chen}, Y., {Liu}, K., {Feng}, S.~W., \& {Xia}, L.~D. 2009,
  \solphys, 258, 129, \dodoi{10.1007/s11207-009-9411-0}

\bibitem[{{Stansby} \& {Horbury}(2018)}]{Stansby2018}
{Stansby}, D., \& {Horbury}, T.~S. 2018, \aap, 613, A62,
  \dodoi{10.1051/0004-6361/201732567}

\bibitem[{{Suess} {et~al.}(2009){Suess}, {Ko}, {von Steiger}, \&
  {Moore}}]{Suess2009}
{Suess}, S.~T., {Ko}, Y.~K., {von Steiger}, R., \& {Moore}, R.~L. 2009, Journal
  of Geophysical Research (Space Physics), 114, A04103,
  \dodoi{10.1029/2008JA013704}

\bibitem[{{Suess} {et~al.}(1996){Suess}, {Wang}, \& {Wu}}]{Suess1996}
{Suess}, S.~T., {Wang}, A.~H., \& {Wu}, S.~T. 1996, \jgr, 101, 19957,
  \dodoi{10.1029/96JA01458}

\bibitem[{{Sánchez-Díaz} {et~al.}(2017{\natexlab{a}}){Sánchez-Díaz},
  {Rouillard}, {Davies}, {Lavraud}, {Pinto}, \& {Kilpua}}]{Sanchez2017b}
{Sánchez-Díaz}, E., {Rouillard}, A.~P., {Davies}, J.~A., {et~al.}
  2017{\natexlab{a}}, \apj, 851, 32, \dodoi{10.3847/1538-4357/aa98e2}

\bibitem[{{Sánchez-Díaz} {et~al.}(2017{\natexlab{b}}){Sánchez-Díaz},
  {Rouillard}, {Davies}, {Lavraud}, {Sheeley}, {Pinto}, {Kilpua}, {Plotnikov},
  \& {Genot}}]{Sanchez2017a}
---. 2017{\natexlab{b}}, \apjl, 835, L7, \dodoi{10.3847/2041-8213/835/1/L7}

\bibitem[{{Thompson} {et~al.}(2003){Thompson}, {Davila}, {Fisher}, {Orwig},
  {Mentzell}, {Hetherington}, {Derro}, {Federline}, {Clark}, {Chen},
  {Tveekrem}, {Martino}, {Novello}, {Wesenberg}, {StCyr}, {Reginald}, {Howard},
  {Mehalick}, {Hersh}, {Newman}, {Thomas}, {Card}, \& {Elmore}}]{Thompson2003}
{Thompson}, W.~T., {Davila}, J.~M., {Fisher}, R.~R., {et~al.} 2003, in Society
  of Photo-Optical Instrumentation Engineers (SPIE) Conference Series, Vol.
  4853, Innovative Telescopes and Instrumentation for Solar Astrophysics, ed.
  S.~L. {Keil} \& S.~V. {Avakyan}, 1--11, \dodoi{10.1117/12.460267}

\bibitem[{{Thomson}(1982)}]{Thomson1982}
{Thomson}, D.~J. 1982, IEEE Proceedings, 70, 1055

\bibitem[{{Tokumaru} {et~al.}(1991){Tokumaru}, {Mori}, {Tanaka}, {Kondo},
  {Takaba}, \& {Koyama}}]{Tokumaru1991}
{Tokumaru}, M., {Mori}, H., {Tanaka}, T., {et~al.} 1991, Journal of
  Geomagnetism and Geoelectricity, 43, 619, \dodoi{10.5636/jgg.43.619}

\bibitem[{{Uchida} {et~al.}(1992){Uchida}, {McAllister}, {Strong}, {Ogawara},
  {Shimizu}, {Matsumoto}, \& {Hudson}}]{Uchida1992}
{Uchida}, Y., {McAllister}, A., {Strong}, K.~T., {et~al.} 1992, \pasj, 44, L155

\bibitem[{{Ventura} {et~al.}(2023){Ventura}, {Antonucci}, {Downs}, {Romano},
  {Susino}, {Spadaro}, {Telloni}, {Guglielmino}, {Capuano}, {Andretta},
  {Landini}, {Jerse}, {Nicolini}, {Pancrazzi}, {Sasso}, {Da Deppo}, {Fineschi},
  {Grimani}, {Heinzel}, {Moses}, {Naletto}, {Romoli}, {Stangalini}, {Teriaca},
  \& {Uslenghi}}]{Ventura2023}
{Ventura}, R., {Antonucci}, E., {Downs}, C., {et~al.} 2023, \aap, 675, A170,
  \dodoi{10.1051/0004-6361/202346623}

\bibitem[{{Viall} \& {Borovsky}(2020)}]{Viall2020}
{Viall}, N.~M., \& {Borovsky}, J.~E. 2020, Journal of Geophysical Research
  (Space Physics), 125, e26005,
  \dodoi{10.1029/2018JA02600510.1002/essoar.10502606.1}

\bibitem[{{Viall} {et~al.}(2021){Viall}, {DeForest}, \& {Kepko}}]{Viall2021}
{Viall}, N.~M., {DeForest}, C.~E., \& {Kepko}, L. 2021, Frontiers in Astronomy
  and Space Sciences, 8, 139, \dodoi{10.3389/fspas.2021.735034}

\bibitem[{{Viall} {et~al.}(2008){Viall}, {Kepko}, \& {Spence}}]{Viall2008}
{Viall}, N.~M., {Kepko}, L., \& {Spence}, H.~E. 2008, Journal of Geophysical
  Research (Space Physics), 113, A07101, \dodoi{10.1029/2007JA012881}

\bibitem[{{Viall} {et~al.}(2009{\natexlab{a}}){Viall}, {Kepko}, \&
  {Spence}}]{Viall2009}
---. 2009{\natexlab{a}}, Journal of Geophysical Research (Space Physics), 114,
  A01201, \dodoi{10.1029/2008JA013334}

\bibitem[{{Viall} {et~al.}(2009{\natexlab{b}}){Viall}, {Spence}, \&
  {Kasper}}]{Viall2009SK}
{Viall}, N.~M., {Spence}, H.~E., \& {Kasper}, J. 2009{\natexlab{b}}, \grl, 36,
  L23102, \dodoi{10.1029/2009GL041191}

\bibitem[{{Viall} {et~al.}(2010){Viall}, {Spence}, {Vourlidas}, \&
  {Howard}}]{Viall2010}
{Viall}, N.~M., {Spence}, H.~E., {Vourlidas}, A., \& {Howard}, R. 2010,
  \solphys, 267, 175, \dodoi{10.1007/s11207-010-9633-1}

\bibitem[{{Viall} \& {Vourlidas}(2015)}]{Viall2015}
{Viall}, N.~M., \& {Vourlidas}, A. 2015, \apj, 807, 176,
  \dodoi{10.1088/0004-637X/807/2/176}

\bibitem[{{Wang} {et~al.}(2012){Wang}, {Grappin}, {Robbrecht}, \&
  {Sheeley}}]{Wang2012}
{Wang}, Y.~M., {Grappin}, R., {Robbrecht}, E., \& {Sheeley}, N.~R., J. 2012,
  \apj, 749, 182, \dodoi{10.1088/0004-637X/749/2/182}

\bibitem[{{Wang} {et~al.}(1999){Wang}, {Sheeley}, {Howard}, {Cyr}, \&
  {Simnett}}]{Wang1999}
{Wang}, Y.~M., {Sheeley}, N.~R., J., {Howard}, R.~A., {Cyr}, O.~C.~S., \&
  {Simnett}, G.~M. 1999, \grl, 26, 1203, \dodoi{10.1029/1999GL900209}

\bibitem[{{Wang} {et~al.}(2000){Wang}, {Sheeley}, {Socker}, {Howard}, \&
  {Rich}}]{Wang2000}
{Wang}, Y.~M., {Sheeley}, N.~R., {Socker}, D.~G., {Howard}, R.~A., \& {Rich},
  N.~B. 2000, \jgr, 105, 25133, \dodoi{10.1029/2000JA000149}

\bibitem[{{Wang} {et~al.}(1998){Wang}, {Sheeley}, {Walters}, {Brueckner},
  {Howard}, {Michels}, {Lamy}, {Schwenn}, \& {Simnett}}]{Wang1998}
{Wang}, Y.~M., {Sheeley}, N.~R., J., {Walters}, J.~H., {et~al.} 1998, \apjl,
  498, L165, \dodoi{10.1086/311321}

\bibitem[{{West} {et~al.}(2023){West}, {Seaton}, {Wexler}, {Raymond}, {Del
  Zanna}, {Rivera}, {Kobelski}, {Chen}, {DeForest}, {Golub}, {Caspi}, {Gilly},
  {Kooi}, {Meyer}, {Alterman}, {Alzate}, {Andretta}, {Auch{\`e}re}, {Banerjee},
  {Berghmans}, {Chamberlin}, {Chitta}, {Downs}, {Giordano}, {Harra},
  {Higginson}, {Howard}, {Kumar}, {Mason}, {Mason}, {Morton}, {Nykyri},
  {Patel}, {Rachmeler}, {Reardon}, {Reeves}, {Savage}, {Thompson}, {Van
  Kooten}, {Viall}, {Vourlidas}, \& {Zhukov}}]{West2023}
{West}, M.~J., {Seaton}, D.~B., {Wexler}, D.~B., {et~al.} 2023, \solphys, 298,
  78, \dodoi{10.1007/s11207-023-02170-1}

\bibitem[{{Wexler} {et~al.}(2020){Wexler}, {Imamura}, {Efimov}, {Song},
  {Lukanina}, {Ando}, {Jensen}, {Vierinen}, \& {Coster}}]{Wexler2020}
{Wexler}, D., {Imamura}, T., {Efimov}, A., {et~al.} 2020, \solphys, 295, 111,
  \dodoi{10.1007/s11207-020-01677-1}

\bibitem[{{Wexler} {et~al.}(2019){Wexler}, {Hollweg}, {Efimov}, {Lukanina},
  {Coster}, {Vierinen}, \& {Jensen}}]{Wexler2019}
{Wexler}, D.~B., {Hollweg}, J.~V., {Efimov}, A.~I., {et~al.} 2019, \apj, 871,
  202, \dodoi{10.3847/1538-4357/aaf6a8}

\bibitem[{{Woo}(1978)}]{Woo1978}
{Woo}, R. 1978, \apj, 219, 727, \dodoi{10.1086/155831}

\bibitem[{{Wuelser} {et~al.}(2004){Wuelser}, {Lemen}, {Tarbell}, {Wolfson},
  {Cannon}, {Carpenter}, {Duncan}, {Gradwohl}, {Meyer}, {Moore}, {Navarro},
  {Pearson}, {Rossi}, {Springer}, {Howard}, {Moses}, {Newmark},
  {Delaboudiniere}, {Artzner}, {Auchere}, {Bougnet}, {Bouyries}, {Bridou},
  {Clotaire}, {Colas}, {Delmotte}, {Jerome}, {Lamare}, {Mercier}, {Mullot},
  {Ravet}, {Song}, {Bothmer}, \& {Deutsch}}]{Wuelser2004}
{Wuelser}, J.-P., {Lemen}, J.~R., {Tarbell}, T.~D., {et~al.} 2004, in Society
  of Photo-Optical Instrumentation Engineers (SPIE) Conference Series, Vol.
  5171, Telescopes and Instrumentation for Solar Astrophysics, ed.
  S.~{Fineschi} \& M.~A. {Gummin}, 111--122, \dodoi{10.1117/12.506877}

\bibitem[{{Zhao} \& {Fisk}(2011)}]{Zhao2011}
{Zhao}, L., \& {Fisk}, L. 2011, \solphys, 274, 379,
  \dodoi{10.1007/s11207-011-9840-4}

\end{thebibliography}
\bibliographystyle{aasjournal}



\end{document}